\documentstyle[11pt,epsf]{article}
\begin{document}

\begin{titlepage}
\samepage{
\setcounter{page}{1}

\rightline{UFIFT--HEP-97-13}
\rightline{ANL-HEP-PR-97-12}
\rightline{May 1997}
\vfill
\begin{center}
 {\Large \bf  Spin Dependent Drell Yan in QCD to $O(\alpha_s^2)$ (I). \\
   (The Non-Singlet Sector) }

\vfill
\vfill
 {\large Sanghyeon Chang$^{*}$\footnote{
        E-mail address: schang@phys.ufl.edu},
        Claudio Corian\`{o}$^{*}$\footnote{Present Address: Theory Group, 
Jefferson Lab, Newport News, 23606 VA, USA.\\
        E-mail address: coriano@jlabs2.jlab.org},   
        R. D. Field$^{*}$\footnote{
        E-mail address: rfield@phys.ufl.edu},
        \\$\,$and$\,$ L. E. Gordon$^{\dagger}$ \footnote 
{E-mail address: gordon@hep.anl.gov}}
\\
\vspace{.12in}
 {\it $^{*}$   Institute for Fundamental Theory, Department of Physics, \\
        University of Florida, Gainesville, FL 32611, 
        USA\footnote{Permanent address.}\\}

\vspace{.075in}

{\it  $^{\dagger}$ Argonne National Laboratory\\
9700 South Cass Av, IL USA\\}
\end{center}
\vfill
\begin{abstract}
A study of the order $\alpha_s^2$ corrections to the 
Drell-Yan ({\it non-singlet}) differential cross section  for incoming states of
arbitrary longitudinal helicities is presented.  The transverse momentum
$distributions$, $q_T$, of the lepton pair are studied and the calculations of 
Ellis, Martinelli, and Petronzio (EMP) are extended to include polarized initial states. 
We use the $\overline{MS}$ scheme and the 
t'Hooft-Veltman regularization for the helicity projectors. 
{}From our results one can obtain the bulk of the totally inclusive 
NNLO cross section for the production of a Drell-Yan pair in the 
non-singlet sector by a simple integration over the virtual photon momentum.  
We show that in the $\overline{MS}$ scheme helicity is not conserved along 
the quark lines, unless 
a finite renormalization is done and one adapts the physical
($\overline{MS}_p$) scheme. This aspect of the calculation is similar to the 
$O(\alpha_s^2)$
polarized production of single and double photons.  Our spin averaged 
{\it unpolarized} differential cross
sections agree with the EMP calculations.
   
\end{abstract}
\smallskip}
\end{titlepage}

\setcounter{footnote}{0}

\def\beq{\begin{equation}}
\def\eeq{\end{equation}}
\def\beqn{\begin{eqnarray}}
\def\eeqn{\end{eqnarray}}

\def\ie{{\it i.e.}}
\def\eg{{\it e.g.}}
\def\half{{\textstyle{1\over 2}}}
\def\nicefrac#1#2{\hbox{${#1\over #2}$}}
\def\third{{\textstyle {1\over3}}}
\def\quarter{{\textstyle {1\over4}}}
\def\m{{\tt -}}

\def\p{{\tt +}}

\def\slash#1{#1\hskip-6pt/\hskip6pt}
\def\slk{\slash{k}}
\def\GeV{\,{\rm GeV}}
\def\TeV{\,{\rm TeV}}
\def\y{\,{\rm y}}

\def\l{\langle}
\def\r{\rangle}

\setcounter{footnote}{0}
\newcommand{\beqa}{\begin{eqnarray}}
\newcommand{\eeqa}{\end{eqnarray}}
\newcommand{\eps}{\epsilon}


\section{Introduction}

In recent years spin physics has become a very active research area in
particle physics.  This is especially true in Deep Inelastic Scattering (DIS), 
due to the puzzling results of the EMC and SMC collaborations on the 
nucleon spin and the experimental activity at HERA. 
Various reviews have appeared in the literature discussing the subject 
in detail and highlighting  
the recent progress made in the area \cite{ans,ramsey,bour,cheng,jacob}.

Along with the possibility of probing the spin structure 
of the hadrons at RHIC and HERA, and possibly at other colliders,
comes the need to investigate systematically the role of subleading effects 
in polarized collisions.  Given the added complexity that spin physics brings 
into the program, it is clear that these new studies will be far more 
involved and will require considerably more effort than in the
past, both on the experimental and on the theoretical fronts. 
It is generally expected that radiative corrections to various QCD  
processes in the case of polarized initial states will be substantial and 
crucial in order to reach a better understanding of the behaviour of the 
polarized parton distributions. In particular, in contrast to the DIS case, 
at hadron colliders the direct gluon coupling will 
allow one to study the size of the longitudinal spin asymmetries of {\it both} quarks
and gluons thereby reaching a more complete 
understanding of the distribution of the spin of the nucleon among its 
constituents.  The Drell-Yan process has received an 
extraordinary amount of attention in the past, both as a way of testing perturbative 
QCD and as a possible mechanism to uncover new gauge interactions at high 
energy.

Leading power factorization theorems predict that high energy scattering 
processes can be written in a factorized form, in which a short distance 
coefficient is convoluted with a non-perturbative universal matrix element. 
At leading twist ({\it or twist 2}) these matrix elements describe 
distributions of 
gluons and quarks in the hadrons, while contributions of higher twist describe
more complex correlations among partons. 
In previous works (see for instance \cite{CG} and refs. therein), 
complete NLO 
analyses of some important polarized processes have been presented. The
results show that radiative corrections are sizeable in many cases and can be 
measured in forthcoming experiments at RHIC and possibly HERA. In this work we 
extend the program outlined in \cite{CG} by analyzing the behaviour of the 
Drell-Yan differential cross section in the case of  {\it longitudinally} 
polarized incoming 
states.  In this work, we present a complete NLO study of the {\it  non-singlet}
contribution to the Drell-Yan production of large transverse momentum lepton 
pairs in polarized 
hadron-hadron collisions,  thereby extending the unpolarized work of Ellis, 
Martinelli, and Petronzio 
\cite{EMP}.  Our calculations are performed in the $\overline{MS}$ scheme. The 
factorization and the renormalization at the perturbative order are performed 
in the same scheme.  In this scheme, the t'Hooft Veltman \cite{thv} 
prescription for 
the regularization of chiral states gives a cross section which is 
related to the EMP result, plus additional helicity non-conserving
contributions.   In the $\overline{MS}$ scheme helicity is not conserved 
along the quark lines.
By a redefinition of the splitting functions one 
finds that the result for the polarized case can be trivially related to the 
unpolarized results. 
However, this redefinition results in a different scheme, with polarized 
splitting functions differing from the unpolarized ones by a finite term. 
This pattern has been already noticed in the case of double and single 
photon production \cite{GV,CG}.
The factorization of the process is shown to occur in both schemes, with 
different structures for the helicity non-conserving contributions coming from the 
real and the virtual diagrams. 

The structure of the real and of the virtual 
corrections are similar to the result of \cite{EMP}, 
but modified by the helicity dependent factors $(1-h)$ and 
$h=4\lambda_1\lambda_2$, where $\lambda_1$ and $\lambda_2$ are the
helicities of the initial state fermions.  This work is mainly concerned with 
the calculation 
of the radiative corrections to the parton level cross section for the 
non-singlet case, while the hadronic part of the calculation and a 
phenomenological 
analysis of our results will be presented elsewhere. An extension of our
calculation to include the contributions from initial state gluons is
now in progress.

This paper is organized as follows. In section 2, we examine the general 
helicity structure of the Drell-Yan process and introduce some of the 
differential cross sections. In section 3, we present the calculation of the 
virtual 
contributions to the differential cross section, while the real contributions 
are presented in section 4. The study of the factorization of the process is  
done in section 4.1, where we present the final structure of the factorized 
differential cross section. Our conclusions and perspectives are contained in 
section 5.

\begin{figure}
\centerline{\epsfbox{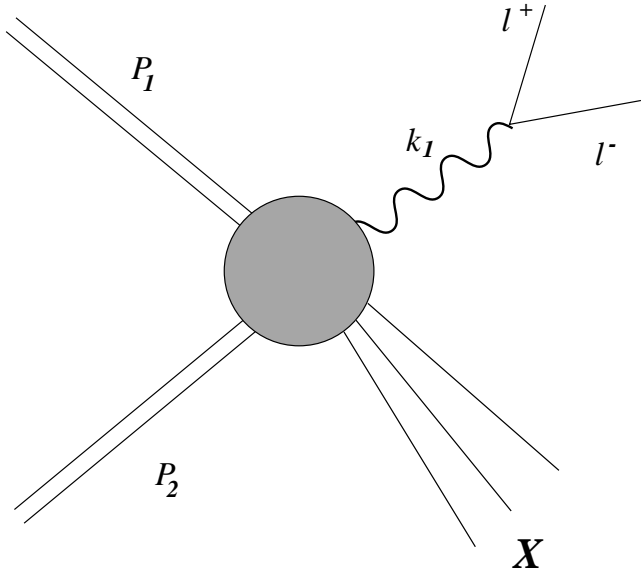}}
\caption{Drell-Yan process}
\end{figure}
Part of the calculations have been performed using FeynCalc \cite{Mertig},
in particular the implementation of the Passarino-Veltman \cite{PV} recursion
procedure for tensor integrals.The presentation of the phase space integration 
can be found in Appendix A, where we describe an extension to the polarized 
case of the previously known methods of integration used in the unpolarized 
case. 

\section{The Drell Yan Process}

Deep Inelastic Scattering (DIS) has provided us with a wealth of information 
on the nucleon structure, however, there are limitations on the
information obtainable from DIS experiments in the study of the spin properties 
of the hadrons.  The first and most important limitation arises because gluons 
couple only to higher order.  Also, distributions of transverse spin decouple.  
For a classification of the various light-cone operators of lowest and 
higher twist see \cite{QS}, \cite{Jaffe1} and references therein.  
The Drell-Yan process is the natural territory to analyze $some$ of the issues 
connected with these important formal developments.  For example, it has been 
observed (see for instance \cite{Jaffe1}) 
that the Drell-Yan production of charged lepton pairs is a 
case where transverse and longitudinal asymmetries are comparable. To leading 
order, the ratio between the transverse and the longitudinal asymmetries, 
$R=A_{TT}/A_{LL}$, is estimated to be of order one. 
This is due to the fact that the gluons don't couple to transverse 
asymmetries and to order $\alpha_s$ only the quark-antiquark channel 
contributes.  In other processes such as inclusive two-jet production 
($p\,\, p\to j\,\,j + X$) or prompt photon production
$\vec{p}\vec{p}\rightarrow \gamma +X$ the gluon channel is available also to 
leading order (in $A_{LL}$), thereby making $R$ small. Therefore, it is very 
important to see how these lowest order predictions are affected by radiative 
corrections. 

Our work is concerned with a  study of the differential cross section for the 
production
of {\it large transverse momentum} lepton pairs, produced by the decay of a 
virtual photon 
of invariant mass $Q^2$.   Since the lepton pair is required to have non-zero 
transverse momentum, $q_T$, the differential cross section starts at order 
$\alpha_s$.  At lowest order (LO),
one gluon ({\it at least}) is required to be in the final state.
The calculation proceeds up to order $\alpha_s^2$ (denoted as NLO). 

\begin{figure}
\epsfxsize=150mm
\centerline{\epsfbox{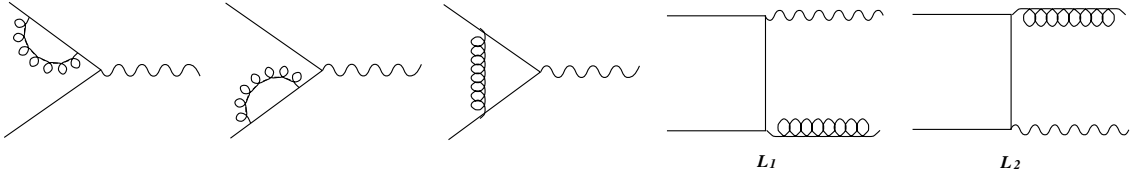}}
\caption{$\alpha_s$ order contributions}
\end{figure}

\subsection{Previous Work}

For the unpolarized case previous work on the process can be found in 
ref.~\cite{EMP} for the non-singlet sector. The total cross section, together 
with a calculation of the $K$ factor and a study of the small-$x$ limit  
has been studied by Van Neerven and collaborators (see \cite{vncol} and 
refs. therein). In the polarized case, the order $\alpha_s$ radiative 
corrections have been originally investigated in \cite{ratcliffe} (total cross 
section).  More recently work has been done to extend this calculation to the 
to the $x_F$-distribution in \cite{geh} to order $\alpha_s$. 
The resummation for the of the differential cross section at small $q_T$ 
of the lepton pair has been studied in the polarized case in \cite{Weber}. The 
integration of our differential cross sections in dimensional regularization 
provides the bulk of the NNLO ($\alpha_s^2$) corrections to the total Drell-Yan
cross section in the non-singlet sector. The remaining contributions, not 
included in this work, can be easily obtained by a simple extension of our 
work, since they amount to finite (in $\epsilon$) terms. However, as we are 
going to see, the structure of the helicity non-conserving contributions do not 
allow us to conclude that at parton level the total cross section is simply
related to the unpolarized result by an overall helicity-dependent factor. 

\subsection{General Structure of the Differential Cross Section}

We begin by examining the contribution to the production of {\it virtual} 
photons with invariant mass, $Q$, in hadron-hadron collisions, 
\beq
H_1 + H_2\to \gamma^* + X,
\label{rdf_x1}
\eeq
from the
$2\to 3$ parton subprocess,
\beq
a+b\to \gamma^*+c+d,
\label{rdf_x2}
\eeq
like those shown in Figs.~4-6.  Here $H_1$ and $H_2$ are incoming hadrons with 
4-momenta, $P_1$ and $P_2$, respectively, and $q$ is the 4-momentum of the 
virtual photon, $\gamma^*$, as shown in Fig.~1.  The 4-momenta of the
incoming two parton $a$ and $b$ are labeled by $p_1$ and $p_2$ respectively, and
the outgoing 4-momenta of partons $c$ and $d$ are labeled by $k_2$ and $k_3$.  
The virtual photon 4-momenta is given by $k_1=q$.  Conservation of energy and 
momentum implies 
that
\beq
p_1+p_2=k_1+k_2+k_3.
\label{rdf_x3}
\eeq
The hadron-hadron process (\ref{rdf_x1}) is described in terms of the
invariants,
\beq
S=(P_1+P_2)^2,\qquad T=(P_1-q)^2,\qquad U=(P_2-q)^2.
\label{rdf_bigs}
\eeq
In addition, it is convenient to define the scaled variables
\beq
x_T=2q_T/W\ {\rm and}\ \tau=Q^2/S,
\eeq
where $W=\sqrt{S}$ is the hadron-hadron
center-of-mass energy and $q_T$ is the transverse momentum of the virtual 
photon with invariant mass, $Q$.  It is also useful to define the following 
two quantities,
\beq
\bar x_1={Q^2-U\over S}=\nicefrac12 e^y\sqrt{x_T^2+4\tau},\qquad
\bar x_2={Q^2-T\over S}=\nicefrac12 e^{-y}\sqrt{x_T^2+4\tau},
\label{rdf_xbar}
\eeq
where $y$ is the rapidity of the virtual photon.
The $2\to 3$ parton subprocess (\ref{rdf_x2}) is described in terms of the 
invariants,
\beq
s=(p_1+p_2)^2,\qquad t=(p_1-q)^2,\qquad u=(p_2-q)^2,
\label{rdf_stu}
\eeq
and
\beq
s_2=s_{23}=(k_2+k_3)^2,
\eeq
where momentum and energy conservation (\ref{rdf_x3}) implies
\beq
s+t+u=Q^2+s_2.
\label{rdf_sum1}
\eeq

In QCD the hadronic cross section is related to the parton subprocess 
according to
\begin{eqnarray}
\lefteqn{d\sigma(H_1+H_2\to \gamma^*+X;W,q_T,y)=} \nonumber \\
&&G_{H_1\to a}(x_1,M^2)dx_1 
G_{H_2\to b}(x_2,M^2)dx_2\left({d\hat\sigma\over dtdu}
(ab\to\gamma^*cd;s,t,u)\right) dtdu,
\label{rdf_eq1}
\end{eqnarray}
where $G_{H_1\to a}(x_1,M^2)dx_1$ is the number of partons of flavor $a$ with 
momentum fraction, $x_1=p_1/P_1$, within hadron $H_1$ at the factorization 
scale $M$.  Similarly, $G_{H_2\to b}(x_2,M^2)dx_2$ is the number of partons of 
flavor $b$ with momentum fraction, $x_2=p_2/P_2$, within hadron $H_2$ at the 
factorization scale $M$.  In the remainder of this paper we will take the 
factorization scale to be $Q$ and evaluate the parton subprocesses at the same 
scale $Q$.  Using (\ref{rdf_bigs}) and (\ref{rdf_stu}) we see that
\begin{eqnarray}
s&=&x_1x_2S,\\
(t-Q^2)&=&x_1(T-Q^2)=-x_1\bar x_2S,\\
(u-Q^2)&=&x_2(U-Q^2)=-x_2\bar x_1S,
\end{eqnarray}
which from (\ref{rdf_sum1}) implies that
\beq
x_1x_2-x_1\bar x_2-x_2\bar x_1=\tau_2-\tau,
\label{rdf_sum2}
\eeq
where $\bar x_1$ and $\bar x_2$ are defined in (\ref{rdf_xbar})
and
\beq
\tau_2=s_2/S.
\eeq
It is now easy to compute the Jacobian
\beq
dx_2dt={s\over4(x_1-\bar x_1)}\ dx_T^2dy,
\eeq
which when inserted into (\ref{rdf_eq1}) and integrating over $x_1$ and
$s_2$ gives
\begin{eqnarray}
\lefteqn{S{d\sigma\over dq_T^2dy}(W,q_T,y)=
\int_{x_1^{min}}^1dx_1\int_0^{s_2^{max}}ds_2\ {1\over (x_1-\bar x_1)}} 
\nonumber \\
&&\ G_{H_1\to a}(x_1,Q^2) 
G_{H_2\to b}(x_2,Q^2)\ s{d\hat\sigma\over dtdu}
(ab\to\gamma^*cd;s,t,u),
\label{rdf_eq2}
\end{eqnarray}
where
\beq
x_2={x_1\bar x_2+\tau_2-\tau\over x_1-\bar x_1},
\eeq
and
\beq
s_2^{max}=A=(\tau-\bar x_1+x_1(1-\bar x_2))S,\qquad
x_1^{min}={\bar x_1-\tau\over 1-\bar x_2}.
\label{rdf_defa}
\eeq
The maximum value of $\tau_2$ arises when
$x_2=1$ in (\ref{rdf_sum2}), while the minimum value of $x_1$ occurs when 
$\tau_2=0$ and $x_2=1$. 

For $2\to 2$ parton subprocesses such as the Born contribution in Fig.~2 one has
\beq
s{d\hat\sigma\over dtdu}(s,t,u)=\delta(s_2)\ s{d\hat\sigma\over dt}
(a+b\to\gamma^*+c;s,t),
\eeq
which when inserted into (\ref{rdf_eq2}) results in
\begin{eqnarray}
\lefteqn{S{d\sigma\over dq_T^2dy}(W,q_T,y)=
\int_{x_1^{min}}^1dx_1\ {1\over (x_1-\bar x_1)}} 
\nonumber \\
&&\ G_{H_1\to a}(x_1,Q^2) 
G_{H_2\to b}(x_2,Q^2)\ s{d\hat\sigma\over dtdu}
(ab\to\gamma^*c;s,t),
\label{rdf_eq3}
\end{eqnarray}
where in this case $s_2=0$ and $s+t+u=Q^2$. Finally, the Drell-Yan differential 
cross section for producing muon pair with transverse momentum $q_T$ and 
invariant mass, $Q$, at rapidity $y$ in the hadron-hadron collision,
\beq
H_1+H_2\to (\gamma^*\to \mu^+\mu^-)+X,
\eeq
at center-of-mass energy, $W$, is given by
\begin{eqnarray}
\lefteqn{S{d\sigma\over dQ^2dq_T^2dy}(H_1H_2\to\mu^+\mu^-+X;W,q_T,y)=} \nonumber \\
&&\left({\alpha\over3\pi Q^2}\right)
S{d\sigma\over dq_T^2dy}(H_1H_2\to\gamma^*+X;W,q_T,y),
\end{eqnarray}
where $\alpha$ is the electromagnetic fine structure constant and
where the virtual photon differential cross section is given by (\ref{rdf_eq2}) 
or (\ref{rdf_eq3}). In the remainder of this paper we will concentrate on the 
virtual photon differential cross section.

\subsection{The Non-Singlet Cross Section}

We define the non-singlet to be the difference of the antihadron-hadron and 
the hadron-hadron differential cross sections as follows:
\begin{eqnarray}
\lefteqn{S{d\sigma_{NS}\over dq_T^2dy}(H+H_2\to\gamma^*+X)=}\nonumber \\
&&S{d\sigma\over dq_T^2dy}(\bar H+H_2\to\gamma^*+X)-
S{d\sigma\over dq_T^2dy}(H+H_2\to\gamma^*+X).
\end{eqnarray}
It is easy to see that
\begin{eqnarray}
\lefteqn{S{d\sigma_{NS}\over dq_T^2dy}(W,q_T,y)=\sum_{f,f'}
\int_{x_1^{min}}^1dx_1\int_0^{s_2^{max}}ds_2\ {1\over (x_1-\bar x_1)}} 
\nonumber \\
&&\ G^{NS}_{H\to q_f}(x_1,Q^2) 
G^{NS}_{H\to q_{f'}}(x_2,Q^2)\ s{d{\hat\sigma}_{NS}\over dtdu}(s,t,u),
\label{rdf_ns1}
\end{eqnarray}
where the non-singlet parton-parton differential cross section is given by
\beq
s{d{\hat\sigma}_{NS}\over dtdu}=
s{d{\hat\sigma}\over dtdu}(\bar q_fq_{f'}\to\gamma^*+X)-
s{d{\hat\sigma}\over dtdu}(q_fq_{f'}\to\gamma^*+X),
\label{rdf_ns2}
\eeq
and the non-singlet structure function is defined according to
\beq
G^{NS}_{H\to q}=G_{H\to q}-G_{H\to \bar q}.  
\eeq
The sum in (\ref{rdf_ns1}) is over the quark flavors $f$ and $f'$.  As we will 
see this sum turns out to be diagonal in flavor space so that one needs only to 
sum over $f=f'$.

We will organize the Feynman diagrams in the same way as was done in 
\cite{EMP}.  At order $\alpha_s$ the Born term comes only from the square of the
two amplitudes in Fig.~2, which we write schematically as
\beq
d{\hat\sigma}_B\sim |L_1+L_2|^2.
\label{rdf_ampb}
\eeq
The Born term contributes only to the diagonal $\bar q_fq_f$ part of the 
non-singlet differential cross section in (\ref{rdf_ns1}) since it is non zero 
only when $f=f'$.

The order $\alpha_s^2$ contributions are separated into the real and the virtual
 diagrams.  The virtual term, $d{\hat\sigma}^{virtual}$ comes from the 
interference between the amplitudes $L_1$ and $L_2$ and the eleven amplitudes 
shown in Fig.~6,
\beq
d{\hat\sigma}^{virtual}\sim 
2Re\left((L_1+L_2)\sum_{i=1}^{11}V_i^*\right),
\label{rdf_ampv}
\eeq
which again contributes only to the diagonal $\bar q_fq_f$ part of the 
non-singlet differential cross section in (\ref{rdf_ns1}).  

The other real diagonal $\bar q_fq_f$ diagrams are
\beq
d{\hat\sigma}^{real}_{\bar q_fq_f}\sim
|\sum_{i=1}^8F_i|^2+|\sum_{i=1}^8G_i|^2,
\eeq
where the amplitudes $F_i$ are shown in Fig.~3 and the amplitudes $G_i$ are 
shown in Fig.~4 with $f=f'$.  The absolute square of the $F_i$ amplitudes can 
be written as follows:
\begin{eqnarray}
\lefteqn{|\sum_{i=1}^8F_i|^2=}\nonumber \\
&&|F_1+F_2|^2+|F_3+F_4|^2+2Re\left(\sum_{i=1}^4F_i\sum_{i=5}^8F^*_i\right)
+|\sum_{i=5}^8F_i|^2,
\end{eqnarray}
where we have used the fact that
\beq
Re\left((F_1+F_2)(F_3+F_4)^*\right)=0.
\eeq
Thus the diagonal real $\bar q_fq_f$ contributions are,
\beq
d{\hat\sigma}^{real}_{\bar q_fq_f}= d{\hat\sigma}_1
+d{\hat\sigma}_2+d{\hat\sigma}_3+d{\hat\sigma}_F,
\label{rdf_nsbar}
\eeq
where
\begin{eqnarray}
d{\hat\sigma}_1&\sim& |F_1+F_2|^2+|\sum_{i=1}^8G_i|^2,\label{rdf_amp1}\\
d{\hat\sigma}_2&\sim& |F_3+F_4|^2,\label{rdf_amp2}\\
d{\hat\sigma}_3&\sim& 2Re\left(\sum_{i=1}^4F_i\sum_{i=5}^8F^*_i\right),\label{rdf_amp3} \\
d{\hat\sigma}_F&\sim& |\sum_{i=5}^8F_i|^2.
\end{eqnarray}

The real diagonal $q_fq_f$ diagrams are
\beq
d{\hat\sigma}^{real}_{q_fq_f}\sim
|\sum_{i=1}^4H_i-\sum_{i=4}^8H_i|^2,
\eeq
where the amplitudes $H_i$ are shown in Fig.~5 and the relative minus sign between 
the direct and exchange diagrams is due to Fermi statistics since $f=f'$. The 
absolute square of the $H_i$ amplitudes can be written as follows:
\begin{eqnarray}
\lefteqn{|\sum_{i=1}^4H_i-\sum_{i=4}^8H_i|^2=} \nonumber \\
&&|\sum_{i=1}^4H_i|^2+|\sum_{i=4}^8H_i|^2
-2Re\left(\sum_{i=1}^4H_i\sum_{i=4}^8H_i^*\right).
\end{eqnarray}
Thus the diagonal real $q_fq_f$ contributions are,
\beq
d{\hat\sigma}^{real}_{q_fq_f}= 
d{\hat\sigma}_H+d{\hat\sigma}_4,
\label{rdf_nsqq}
\eeq
where
\begin{eqnarray}
d{\hat\sigma}_H&\sim& |\sum_{i=1}^4H_i|^2+|\sum_{i=4}^8H_i|^2,\\
d{\hat\sigma}_4&\sim& -2Re\left(\sum_{i=1}^4H_i\sum_{i=4}^8H_i^*\right)\label{rdf_amp4}.
\end{eqnarray}
The diagonal ($f=f'$) real part of the non-singlet cross section in (\ref{rdf_ns2}) is given by subtracting
(\ref{rdf_nsqq}) from (\ref{rdf_nsbar}) as follows:
\beq
d{\hat\sigma}^{real}_{NS}=d{\hat\sigma}_1
+d{\hat\sigma}_2+d{\hat\sigma}_3-d{\hat\sigma}_4,
\label{rdf_ns3}
\eeq
where we have used the fact that $d{\hat\sigma}_F=d{\hat\sigma}_H$ which 
arises due to
\beq
\int_{PS_3} |\sum_{i=5}^8F_i|^2
=\nicefrac12 \int_{PS_3} |\sum_{i=1}^4H_i|^2
+\nicefrac12 \int_{PS_3}|\sum_{i=4}^8H_i|^2.
\eeq
When integrating over the two identical particles $k_3$ and $k_4$ over the 
phase space $PS_3$ in (\ref{ps1}) an extra statistical factor of $1/2$ must be 
inserted so as not to double count.  
Thus, the diagonal ($f=f'$) part of the complete non-singlet cross section 
in (\ref{rdf_ns2}) ({\it real} plus {\it virtual}) is given by
\beq
d{\hat\sigma}_{NS}=d{\tilde\sigma}_1
+d{\hat\sigma}_2+d{\hat\sigma}_3-d{\hat\sigma}_4,
\label{rdf_ns4}
\eeq
where
\beq
d{\tilde\sigma}_1=d{\hat\sigma}^{fact}_1+d{\hat\sigma}^{virtual},
\eeq
and where $d{\hat\sigma}^{fact}_1$ is the ``factorized" cross section given by
\beq
d{\hat\sigma}^{fact}_1=d{\hat\sigma}_1+d{\hat\sigma}^{counter}_1.
\label{rdf_counter}
\eeq
It is necessary to add a ``counterterm" cross section 
$d{\hat\sigma}^{counter}_1$ to $d{\hat\sigma}_1$ to cancel the initial state 
collinear singularities.

The off-diagonal real $\bar q_fq_{f'}$ diagrams are
\beq
d{\hat\sigma}^{real}_{\bar q_fq_{f'}}\sim
|\sum_{i=1}^4F_i|^2,
\eeq
where the amplitudes $F_i$ are shown in Fig.~3 and where $f\neq f'$.
Similarly, the off-diagonal real diagonal $q_fq_{f'}$ diagrams are
\beq
d{\hat\sigma}^{real}_{q_fq_{f'}}\sim
|\sum_{i=1}^4H_i|^2,
\eeq
where the amplitudes $H_i$ are shown in Fig.~5.  In this case,
\beq
|\sum_{i=1}^4F_i|^2=|\sum_{i=1}^4H_i|^2,
\eeq
so that these two contributions cancel when forming the non-singlet cross 
section in (\ref{rdf_ns2}).  Thus to order $\alpha_s^2$ (\ref{rdf_ns1})
becomes
\begin{eqnarray}
\lefteqn{S{d\sigma_{NS}\over dq_T^2dy}(W,q_T,y)=\sum_{f}
\int_{x_1^{min}}^1dx_1\int_0^{s_2^{max}}ds_2\ {1\over (x_1-\bar x_1)}} 
\nonumber \\
&&\ G^{NS}_{H\to q_f}(x_1,Q^2) 
G^{NS}_{H\to q_f}(x_2,Q^2)\ s{d{\hat\sigma}_{NS}\over dtdu}(s,t,u),
\label{rdf_nsf1}
\end{eqnarray}
where the sum is now only over the quark flavor $f$ and where
\beq
s{d{\hat\sigma}_{NS}\over dtdu}=
s{d{\hat\sigma}_{B}\over dtdu}+
s{d{\tilde\sigma}_{1}\over dtdu}+
s{d{\hat\sigma}_{2}\over dtdu}+
s{d{\hat\sigma}_{3}\over dtdu}-
s{d{\hat\sigma}_{4}\over dtdu}.
\label{rdf_nsf2}
\eeq

\subsection{The Helicity Dependence of the Born Term}

The lowest order ({\it non-singlet}) contributions to the large 
transverse momentum production of virtual photons  (see Fig.~2) with 
invariant mass $q^2=Q^2$ arise from the two, $q+\bar{q}\to 
\gamma^*+g$,  ``Born" amplitudes $L_1$ and $L_2$.  We refer 
to these two diagrams as the {\it direct} and the {\it exchange} 
(or crossed) amplitudes, respectively. The absolute square of the sum of these 
amplitudes form the Born differential cross section in (\ref{rdf_ampb}).  
In Fig.~2 we have also generically illustrated the expansion of the amplitude 
which appears in $d\sigma/d\,Q^2$ up to order $\alpha_s$. Notice that the 
quark form factor contributions in Fig.2
(and the related $O(\alpha_s^2)$ corrections, not included in the picture) 
do not appear in the study of the cross section, 
$d\sigma/dq^2_T d\,y$, for transverse momentum, $q_T$, greater than zero. 
The two-loop $O(\alpha_s^2)$ corrections 
to the quark form factor can be added in the study of 
the total cross section, $d\sigma/d\,Q^2$, by interfering the 2-loop on-shell 
quark form factor of ref.~\cite{formfac} with the lowest order 
$q\bar{q}\gamma^*$ annihilation channel and by using helicity projectors 
for the initial quark states.

We calculate the spin dependence of the cross section by using the helicity 
projectors, $P_{\pm}={1\over 2}(1 \pm \gamma_5)$, which project out 
the helicity states of an initial state quark and antiquark, 
respectively, as follows:
\beq
u(p_1,h_1)={1\over 2}(1+h_1\gamma_5)u(p_1),\qquad
\bar{v}(p_2,h_2)={1\over 2}\bar{v}(p_2)(1+h_2 
\gamma_5),
\eeq
where $h_1=\pm 1=2\lambda_1$ corresponds to quark helicity 
$\lambda_1=\pm {1\over 2}$, and $h_2=\pm 1=2\lambda_2$ corresponds to antiquark 
helicity $\lambda_2=\pm {1\over 2}$.

The squares of the direct amplitude $L_1$ and exchange amplitude 
$L_2$ in Fig.~2 in $N=4-2\epsilon $ dimensions are given by
\beqn
&& M_{dd}(h_1,h_2)=e_f^2g^2g_s^2
{C_F\over N_c}
{2u\over t}\left( (1-\eps)^2-h_1h_2(1+\eps)^2\right),\nonumber \\
&& M_{cc}(h_1,h_2)=e_f^2g^2g_s^2
{C_F\over N_c}
{2t\over u}\left( (1-\eps)^2-h_1 h_2(1+\eps)^2\right).
\eeqn
where $\alpha=g^2/4\pi$ is the fine structure constant, and 
$e_f$ the charge of the quark.  The quantity $C_F/N_c$ is the 
color factor, and $\alpha_s=g_s^2/4\pi$ is the QCD strong 
coupling constant. 
We pause here
to note that in our NLO calculation $\alpha_s\to\alpha_s(\mu^2)$ is the
running coupling constant renormalized at the scale $\mu^2$ in the
$\overline{MS}$ scheme satisfying the renormalization group equation
\begin{equation}
\mu^2\frac{d\alpha_s}{d\mu^2}=-\alpha_s\left[
\beta_0\frac{\alpha_s}{4\pi}+\beta_1\left(
\frac{\alpha_s}{4\pi}\right)^2+O(\alpha_s^3)\right]
\end{equation}
with
\begin{eqnarray}
\beta_0&=&\frac{11 N_C}{3}-\frac{2 N_F}{3} \nonumber \\
\beta_1&=&\frac{34 N_C^2}{3}-\frac{10 N_C N_F}{3}-2 C_F N_F
\end{eqnarray}
and with $N_F$ the number of active quark flavors.
The interference term is more complicated 
and is given by,
\beqn
&& 2M_{dc}(h_1,h_2)= \nonumber \\
&& e_f^2g^2g_s^2
{C_F\over N_c}
{4\over tu}\left[ (1-\eps)(Q^2s-\eps tu)\right.
\left.-h_1h_2(1+\eps)(Q^2s+\eps tu)-2h_1h_2\eps 
tu\right].
\eeqn
The sum of the two Born amplitudes squared is 
\beqn
&& |M_{B}(h_1,h_2)|^2=|M_{B}(h)|^2=M_{dd}(h)+2M_{dc}(h
)+M_{cc}(h)\nonumber \\
&& =e_f^2g^2g_s^2
{C_F\over N_c}
{2\over tu}\left[(1-\eps)\left(2Q^2s+(1-\eps)(t^2+u^2)-2\eps 
tu\right)\right.\nonumber \\
&&\left.-h(1+\eps)\left(2Q^2s+(1+\eps)(t^2+u^2)+2\eps tu\right) 
-4h\eps tu\right],
\eeqn
and depends only on the product $h=h_1h_2$.  

To any order in perturbation theory we can write,
\beq
|M(h_1,h_2)|^2=|M(h)|^2=(1-h)|\overline{M}|^2+h|M_{++}|^2 
=|\overline{M}|^2+h\Delta |M|^2,
\eeq
where $h=h_1h_2=4\lambda_1\lambda_2$ and where
\beq
|\overline{M}|^2={1\over4} 
|\sum_{h_1,h_2}M(h_1,h_2)|^2={1\over4}
\sum_{h_1,h_2}|M(h_1,h_2)|^2,
\eeq
is the spin averaged ({\it unpolarized}) amplitude squared.  
Furthermore,
\beq
|M_{--}|^2=|M_{++}|^2\ {\rm and}\  |M_{-+}|^2=|M_{+-}|^2,
\eeq
so that
\beq
|\overline{M}|^2={1\over2} \left( |M_{++}|^2+|M_{+-}|^2\right).
\eeq
The spin asymmetry, $\Delta|M|^2$, is defined according to
\beq
\Delta|M|^2={1\over2}\left( |M_{++}|^2-
|M_{+-}|^2\right)
=|M_{++}|^2-|\overline{M}|^2.
\eeq

The spin averaged ({\it unpolarized}) amplitude squared is 
determined from $|M(h)|^2$ by setting $h=0$ with 
$ \Delta|M|^2$ the coefficient of $h$.  For the Born term, this results 
in
\beq
|\overline{M}|^2=e_f^2g^2g_s^2
{C_F\over N_c}
{2\over tu}\left[(1-\eps)\left(2Q^2s
+(1-\eps)(t^2+u^2)-2\eps tu\right)\right],
\eeq
and
\beq
\Delta|M|^2=-e_f^2g^2g_s^2
{C_F\over N_c}
{2\over tu}\left[(1+\eps)\left(2Q^2s +(1+\eps)(t^2+u^2)+2\eps 
tu\right)+4\eps tu\right].
\eeq
Adding these two terms yields
\beq
|M_{++}|^2=-e_f^2g^2g_s^2
{C_F\over N_c}
{8\eps\over tu}\left[Q^2s+(t+u)^2\right],
\eeq
which is proportional to $\eps$ and vanishes in the limit 
$\eps\rightarrow 0$.  Since at the Born level there are no 
${1\over\eps}$ singularities that might combine with this term 
to yield a finite contribution, in the limit $\eps\rightarrow 0$,
\beq
|M_{++}|^2=0\ {\rm and}\ \Delta |M|^2=-|\overline{M}|^2.
\label{spin1}
\eeq
For the $q+\bar{q}\to \gamma^*+g$ subprocess the condition that
$|M_{++}|^2=0$ means that the quark helicity is 
maintained (does not flip) in the collision. The incoming
quark line with helicity $\pm {1\over 2}$ turns around and becomes an
incoming antiquark with helicity $\mp {1\over 2}$.  We refer to this as 
``helicity conservation".  We see that at the Born level helicity is
conserved in the limit $\eps\rightarrow 0$.

In $N=4-2\eps$ dimensions  the differential cross section is 
related to the $2\to 2$ invariant amplitude according to
\beq
s{d\hat\sigma\over dt}(s,t,h)={1\over 16\pi s}
\left({4\pi s\over tu}\right)^\eps{1\over\Gamma(1-\eps)}
|M(h)|^2,
\eeq
which can be written as
\beq
s{d\hat\sigma\over dt}(s,t,h)=(1-h)\ s{d\hat\sigma^\Sigma\over dt}(s,t)+
h\ s{d\hat\sigma_{++}\over dt}(s,t),
\eeq
or as
\beq
s{d\hat\sigma\over dt}(s,t,h)=s{d\hat\sigma^\Sigma\over dt}(s,t)+
h\ s{d\hat\sigma_{LL}\over dt}(s,t),
\eeq
where $sd\hat\sigma^\Sigma/dt$ is the unpolarized cross section
\beq
s{d\hat\sigma^\Sigma\over dt}={1\over2}\left(s{d\hat\sigma_{++}\over 
dt}+s{d\hat\sigma_{+-}\over dt}\right).
\label{rdf_sigsig}
\eeq
and
\beq
s{d\hat\sigma_{LL}\over dt}={1\over2}\left(s{d\hat\sigma_{++}\over 
dt}-s{d\hat\sigma_{+-}\over dt}\right).
\label{rdf_sigll}
\eeq
At the Born level we have,
\beqn
&& s{d\hat\sigma_B^\Sigma\over dt}(s,t)=e_f^2K_2{\alpha_s\over 
s}T_B(Q^2,u,t), \nonumber \\
&& s{d\hat\sigma_B^{++}\over dt}(s,t)=-e_f^2K_2{\alpha_s\over 
s}\epsilon A_B(Q^2,u,t),
\label{rdf_born1}
\eeqn
where
\beqn
&& T_B(Q^2,u,t)={2\over tu}\left[(1-\eps)\left(2Q^2s
+(1-\eps)(t^2+u^2)-2\eps tu\right)\right] \nonumber \\
&& =2(1-\eps)\left[(1-\eps)\left({u\over t}+{t\over u}\right)+
{2Q^2(Q^2-u-t)\over ut}-2\eps\right],
\label{defTB}
\eeqn
and
\beq
A_B(Q^2,u,t)={8\over tu}\left[Q^2s+(t+u)^2\right],
\label{defAB}
\eeq
where $K_2$ is defined by
\beq
K_2=\pi\alpha{C_F\over N_c}
{1\over\Gamma(1-\eps)}
\left({4\pi\mu^2\over Q^2}\right)^\eps\left({sQ^2\over 
tu}\right)^\eps,
\eeq
where we have rescaled, $\alpha_s\to\alpha_s(\mu^2)^\eps$, so 
that it remains dimensionless in $N=4-2\eps$ dimensions. 
At this order of perturbation theory, we have
\beq
s{d\hat\sigma_B\over dt}(s,t,h)=e_f^2K_2{\alpha_s\over 
s}\left[(1-h)T_B(Q^2,u,t)-h\epsilon A_B(Q^2,u,t)\right].
\label{bornddt}
\eeq
As we saw earlier, in the limit $\epsilon\to 0$ helicity is
conserved so that
\beq
s{d\hat\sigma_B\over dt}(s,t,h)=(1-h)s{d\hat\sigma_B^\Sigma\over dt}(s,t),
\label{spin2}
\eeq
which implies that 
\beq
s{d\hat\sigma_B^{++}\over dt}(s,t)=0\ {\rm and}\ s{d\hat\sigma_B^{LL}\over dt}(s,t)=-s{d\hat\sigma_B^\Sigma\over dt}(s,t),
\eeq
so that to leading order,
\beq
\hat A_{LL}={d\hat\sigma_{LL}\over dt}\bigg/{d\hat\sigma^\Sigma\over dt}=-1.
\label{rdf_allborn}
\eeq

To connect with the notation of ref.~\cite{EMP}, we note that
\beq
K_2T_B(Q^2,u,t)=KT_0(Q^2,u,t),
\eeq
where $K$ and $T_0$ are define in ref.~\cite{EMP} as
\beq
K=2\pi\alpha{C_F\over N_c}
{(1-\eps)\over\Gamma(1-\eps)}
\left({4\pi\mu^2\over Q^2}\right)^\eps\left({sQ^2\over 
tu}\right)^\eps,
\eeq
and
\beq
T_0(Q^2,u,t)=\left[(1-\eps)\left({u\over t}+{t\over u}\right)+
{2Q^2(Q^2-u-t)\over ut}-2\eps\right].
\eeq

\subsection{General Structure of the Helicity Dependent Cross Section}

To order $\alpha_s^2$ the non-singlet helicity dependent Drell Yan cross 
section can be written according to (\ref{rdf_nsf1}).  Namely,
\begin{eqnarray}
\lefteqn{S{d\sigma_{NS}\over dq_T^2dy}(W,q_T,y,\Lambda_1,\Lambda_2)=
\sum_{f}\sum_{\lambda_1,\lambda_2}
\int_{x_1^{min}}^1dx_1\int_0^{s_2^{max}}ds_2\ {1\over (x_1-\bar x_1)}} 
\nonumber \\
&&\ G^{NS}_{H[\Lambda_1]\to q_f[\lambda_1]}(x_1,Q^2) 
G^{NS}_{H[\Lambda_2]\to q_f[\lambda_2]}(x_2,Q^2)\ s{d{\hat\sigma}_{NS}\over dtdu}(s,t,u,h),
\label{rdf_nsfh1}
\end{eqnarray}
where $\Lambda_1$ and $\Lambda_2$ are the helicities of the initial two 
hadrons, respectively, and $h=4\lambda_1\lambda_2$ and 
where the sum is now over the quark flavor $f$ and the parton helicities 
$\lambda_1$ and $\lambda_2$.  The non-singlet polarized structure functions 
are given by
\beq
G^{NS}_{H[\Lambda]\to q_f[\lambda]}=G_{H[\Lambda]\to q_f[\lambda]}
-G_{H[\Lambda]\to \bar q_f[\lambda]}.  
\eeq
The parton-parton non-singlet differential can be written as a function of $h$ as follows:
\begin{eqnarray}
\lefteqn{s{d{\hat\sigma}_{NS}\over dtdu}(s,t,u,h)=}\nonumber \\
&&s{d{\hat\sigma}_{B}\over dtdu}(h)+
s{d{\tilde\sigma}_{1}\over dtdu}(h)+
s{d{\hat\sigma}_{2}\over dtdu}(h)+
s{d{\hat\sigma}_{3}\over dtdu}(h)-
s{d{\hat\sigma}_{4}\over dtdu}(h).
\label{rdf_nsfh2}
\end{eqnarray}
{}From (\ref{spin2}) we see helicity is conserved for the Born term so that
\beq
s{d\hat\sigma_{B}\over dtdu}(s,t,u,h)=
(1-h)\delta(s_2)\ s{d\hat\sigma_B^\Sigma\over dt}(s,t),
\label{rdf_born2}
\eeq
where $d\hat\sigma_B^\Sigma$ is the unpolarized cross section in (\ref{rdf_born1}). 
Helicity is also conserved for $d\hat\sigma_2$ defined in (\ref{rdf_amp2}) 
and for $d\hat\sigma_3$ defined in (\ref{rdf_amp3}) so that
\beq
s{d\hat\sigma_{2}\over dtdu}(s,t,u,h)=
(1-h)\ s{d\hat\sigma_2^\Sigma\over dtdu}(s,t,u),
\label{rdf_sig2h}
\eeq
and
\beq
s{d\hat\sigma_{3}\over dtdu}(s,t,u,h)=
(1-h)\ s{d\hat\sigma_3^\Sigma\over dtdu}(s,t,u).
\label{rdf_sig3h}
\eeq
The Born term and these two cross sections have the property that
$d\hat\sigma_(++)=0$ as expected from helicity conservation.
The term $d\hat\sigma_4$ defined in (\ref{rdf_amp4}) is different.  
Here helicity conservation implies that $d\hat\sigma_(+-)=0$ and we have
\beq
s{d\hat\sigma_{4}\over dtdu}(s,t,u,h)=
(1+h)\ s{d\hat\sigma_4^\Sigma\over dtdu}(s,t,u).
\label{rdf_sig4h}
\eeq
The term $d\tilde\sigma_1$ is complicated and whether helicity is conserved or 
not depends on how one does the factorization in (\ref{rdf_counter}).  We 
write the helicity structure as follows:
\beq
s{d\tilde\sigma_{1}\over dtdu}(s,t,u,h)=
(1-h)\ s{d\tilde\sigma_1^\Sigma\over dtdu}(s,t,u)
+h\ s{d\hat\sigma^{hat}_0\over dtdu}(s,t,u),
\label{rdf_sig1h}
\eeq
where $d\hat\sigma^{hat}_0$ is a regularization scheme dependent helicity 
non-conserving piece and 
$d\tilde\sigma_1^\Sigma$ is the unpolarized spin-averaged cross section.  

To order $\alpha_s^2$ the non-singlet parton level spin asymmetry defined in (\ref{rdf_sigll}) becomes
\begin{eqnarray}
s{d{\hat\sigma}^{LL}_{NS}\over dtdu}(s,t,u)=
-s{d{\hat\sigma}^\Sigma_{NS}\over dtdu}+
s{d{\hat\sigma}^{hat}_0\over dtdu}-
2s{d{\hat\sigma}^\Sigma_{4}\over dtdu},
\label{rdf_nspin1}
\end{eqnarray}
where $d{\hat\sigma}^\Sigma_{NS}$ is the non-singlet unpolarized spin 
averaged cross section given by
\beq
s{d{\hat\sigma}^\Sigma_{NS}\over dtdu}(s,t,u)=
s{d{\hat\sigma}_{B}^\Sigma\over dtdu}+
s{d{\tilde\sigma}_{1}^\Sigma\over dtdu}+
s{d{\hat\sigma}_{2}^\Sigma\over dtdu}+
s{d{\hat\sigma}_{3}^\Sigma\over dtdu}-
s{d{\hat\sigma}_{4}^\Sigma\over dtdu}.
\label{rdf_nspin2}
\eeq
At the parton level it is no longer true that $\hat A_{LL}=-1$ as was the 
case for the Born term in (\ref{rdf_allborn}).

\section{Virtual Diagram Contributions}

The list of the diagrams with virtual corrections contributing to the 
non-singlet sector  is defined in (\ref{rdf_ampv}) and shown in Fig.~6. 
We have omitted all the self-energy insertions of quarks and gluons and the 
ghost contributions. 

Our calculations are performed in the $\overline{MS}$ scheme using dimensional 
regularization to regulate both the ultraviolet and the infrared singularities. 
We remove the ultraviolet singularities in the relevant subdiagrams by off 
shell regularization. Then by sending on shell the initial state quarks and 
the final state gluon, we encounter singularities in the form of double poles 
and single poles in $\epsilon = 2- n/2 $. The renormalization of the diagrams 
is enforced in such a way to get straightforwardly final results free of 
UV singularities by symbolic manipulation. We briefly comment here on how this 
is achieved in our case, although a more detailed discussion of the method 
will be presented elsewhere. 
As a first step, we start with the Passarino-Veltman reduction of 
the tensor contributions and generate the expression of the invariant 
amplitudes of the reduction. In our case only one external line 
(the photon ) is massive. The calculation is ambiguous because 
of the presence of massless partons and, in principle, one has to be 
particularly careful to differentiate between the UV and the IR origin of all 
the singularities, which in our approach happen to be identified 
($\epsilon_{UV}=\epsilon_{IR}$ in Dimensional Regularization).
Therefore, the expressions of the invariant amplitudes in the Passarino Veltman 
reduction  are, at this stage, unrenormalized. 

At the second step, we renormalize 
the scalar diagrams which appear in the result by removing the UV poles, 
(since these are removed by coupling constant renormalization) and then 
switch $\epsilon\to -\epsilon$ in the final result. This allows us to at this 
stage interpret the left over singularities as IR singularities. 
Even at this stage the procedure is not complete. One needs a prescription to 
handle the massless tadpoles (self energies at zero momentum) in a consistent 
way. The Passarino Veltman procedure in the massless case, in fact, suffers 
from ambiguities of this type, since massless tadpoles are intrinsically 
ill-defined (being scaleless). In dimensional regularization they are 
usually set to vanish. However, since we have identified UV and 
IR singularities in our algorithm, this last step is also ambiguous. 

At the third step we impose two prescriptions which allow us 
to handle consistently way all the tadpoles and all the massive self energy 
contributions.
These include :
\begin{enumerate} 
\item isolated tadpoles (B(0)) and ``linear" tadpoles (or $\epsilon B(0)$ 
terms). 
\item massive self energy contributions 
($B(q^2)$ and $\epsilon B(q^2)$).
\end{enumerate}
In (2) we implement off-shell renormalization, while in (1) we reinterpret all 
the scaleless contributions as the massless limit of the off-shell ones. 
After step 3, we obtain the 
correct renormalized expressions of all the Passarino Veltman coefficients for 
all the one-loop amplitude. The method works at one loop for any 
$n$-point function and is extremely convenient in practical applications 
since it allows us to forget about the nature 
of the singularities after $IR/UV$ identification. The symbolic implementation 
of the method is also quite straightforward and will be illustrated elsewhere 
\cite{CCG}.

The one-loop order result for the cross section for the process
$q+\bar{q} \rightarrow \gamma^* + g$
is given by
\begin{eqnarray}
&&s{d\hat\sigma^{\rm virtual}\over dtdu}(s,t,u,h) = \nonumber \\
&&e^2_f K_2\frac{\alpha_s}{s}\delta(s+t+u-Q^2)
\left\{ \left((1-h)T_B-h\epsilon A_B\right) 
\left[ 1 - \frac{\alpha_s}{2\pi}\frac{\Gamma(1-\epsilon)}{\Gamma
(1- 2 \epsilon)}\left(\frac{4\pi \mu^2}{Q^2}\right)^{\epsilon}\right.\right.
\nonumber \\
 &  & \left.\times\left(\frac{2 C_F + N_C}{\epsilon^2} 
+ \frac{1}{\epsilon}
\left(3 C_F - 2 C_F \ln \frac{s}{Q^2} + \frac{11}{6} N_C + N_C \ln
\frac{sQ^2}{ut} -\frac{1}{3}N_F\right)\right)\right]
\nonumber \\
&&+\frac{\alpha}{2\pi}(1-h)\left[\pi^2(4C_F + N_C)\frac{ 2Q^2 s + t^2 + u^2}{3tu}
-2(2C_F- N_C)\frac{Q^2 (t^2 + u^2)}{tu(t +  u)}  \right.
\nonumber \\ &&
 -2 C_F  
\left(\frac{8(2Q^2 s +t^2+u^2)}{tu}  -\frac{Q^4 s(t+u)}{t u
(Q^2-u)(Q^2-t)} -\frac{t^2+u^2}{(Q^2-u)(Q^2-t)}\right)   
\nonumber \\ &&
  -2
 \left( Li_2\left(\frac{t}{t-Q^2}\right) +\frac{1}{2}\ln^2\left(1 - \frac{Q^2}{t}\right) 
\right)
\left(N_C \frac{2 s+ t}{u} + 
     2 C_F \frac{s^2+(s+u)^2}{t u}\right) 
\nonumber \\ &&
- 2
  \left(Li_2\left(\frac{u}{u-Q^2}\right) +\frac{1}{2}\ln^2\left(1 - \frac{Q^2}{u}\right)
\right)
\left(N_C \frac{2 s+ u}{t} 
     +2 C_F \frac{s^2+(s+t)^2}{t u}\right) 
\nonumber \\ &&
+ 2 (2 C_F - N_C)\left(
Li_2\left(-\frac{t + u}{s}\right)
 \frac{2Q^2 s +u^2+t^2+2s^2}{t u} \right.
\nonumber \\ &&
+ \left(2\ln\left(\frac{s}{Q^2}\right) 
 \frac{Q^4 - (t+u)^2 }{(t + u)^2} 
 + \ln^2\left(\frac{s}{Q^2}\right)\frac{s^2}{t u} \right)
\nonumber \\ &&
-  \left(\ln\left(\frac{|t|}{Q^2}\right)\ln\left(\frac{s}{Q^2}\right) 
- \frac{1}{2} \ln^2\left(\frac{|t|}{Q^2}\right) \right)
     \frac{s^2+(s+u)^2}{t u} 
\nonumber \\ &&
\left. - \left(\ln\left(\frac{s}{Q^2}\right) \ln\left(\frac{|u|}{Q^2}\right) 
- \frac{1}{2} \ln^2\left(\frac{|u|}{Q^2}\right) \right)
     \frac{s^2+(s+t)^2}{t u} \right)
\nonumber \\ &&
+2 \ln\left(\frac{|u|}{Q^2}\right)\left( C_F\frac{4 Q^2 s -2 s u + t u }
{(Q^2 - u)^2} + N_C \frac{u}{Q^2 - u} \right)
\nonumber \\ &&
+ 2 \ln\left(\frac{|t|}{Q^2}\right) \left(C_F \frac{4Q^2 s-2 s t+t u
 }{(Q^2 - t)^2 } + N_C  \frac{t}{Q^2 - t} \right) 
\nonumber \\ &&
\left.\left.-2\ln\left(\frac{|t|}{Q^2}\right)\ln\left(\frac{|u|}{Q^2}\right) N_C
\frac{2 Q^2 s+t^2+u^2 }{ t u} 
\right]\right\},
\label{mainr}
\end{eqnarray}
where $h=h_1h_2=4\lambda_1\lambda_2$.
This result agrees with the ({\it unpolarized}) spin-averaged ($h\to 0$) 
result given in 
\cite{EMP}, although its structure is more simplified since we have performed 
some additional manipulations on the finite contributions.
Therefore, we confirm the result of \cite{EMP}, which had been used 
before by other authors \cite{Gonsalves, Arnold}, but never checked before.  
The quantities $T_B$ and $A_B$ appear in at the Born level and are given by 
equations (\ref{defTB}) and (\ref{defAB}), respectively. Notice that the double 
poles and single pole structure (in the $h\to 0$ limit) automatically 
reproduces the singularities of the virtual contributions of 
ref.~\cite{EMP}.  In particular, the double poles in $\epsilon$ correctly 
multiply the two-to-two Born contribution $T_B$, which is the 
Born level unpolarized result. 
The presence of the $h\epsilon A_B$ term in (\ref{mainr}) 
implies that in the t'Hooft-Veltman \cite{thv,bm} scheme the virtual 
corrections by themselves do not conserve helicity (\ie\ they do not satisfy 
the equation (\ref{spin2})).  In particular, the finite part of
$d\hat\sigma_{++}/dt$ in the limit $\epsilon\to 0$ for the virtual corrections is 
given by 
\beqn
&&s{d\hat\sigma_{++}^{\rm virtual}\over dt}(s,t) = \nonumber \\
&&-e^2_f K_2\frac{\alpha_s^2}{2\pi s}
A_B(s,t)\left(3 C_F - 2 C_F \ln \frac{s}{Q^2} + \frac{11}{6} N_C + N_C \ln
\frac{sQ^2}{ut} -\frac{1}{3}N_F\right),
\label{plusplus}
\eeqn
where $s+t+u=Q^2$.
In some regularization schemes, such as those enforcing an anticommuting 
$\gamma_5$ in $n$ dimensions,  helicity would be conserved at parton level. 
In the t'Hooft-Veltman \cite{thv,bm} scheme helicity is not manifestly conserved.

\section{The Real Emission Processes}

{}From (\ref{rdf_nsf1}) and (\ref{rdf_nsf2}) we only need to consider real 
contributions that are
diagonal in flavor space. The diagonal ($f=f'$) real part of the non-singlet 
cross section in (\ref{rdf_ns2}) is given in (\ref{rdf_ns3}) by
\beq
d{\hat\sigma}^{real}_{NS}=d{\hat\sigma}_1
+d{\hat\sigma}_2+d{\hat\sigma}_3-d{\hat\sigma}_4,
\eeq
where 
\beq
d{\hat\sigma}_1\sim  |F_1+F_2|^2+|\sum_{i=1}^8G_i|^2,
\eeq
and where $d{\hat\sigma}_2$, $d{\hat\sigma}_3$, and $d{\hat\sigma}_4$, are defined
in (\ref{rdf_amp2}),  (\ref{rdf_amp3}), and (\ref{rdf_amp4}), respectively.
The cross section $d\hat\sigma_1$ requires factorization,
\beq
d{\hat\sigma}^{fact}_1=d{\hat\sigma}_1+d{\hat\sigma}^{counter}_1,
\eeq
where the ``counterterm", $d{\hat\sigma}^{counter}_1$ to $d{\hat\sigma}_1$,  
cancels the initial state collinear singularities.  The factored cross section 
is then added to the virtual term to give,
\beq
d{\tilde\sigma}_1=d{\hat\sigma}^{fact}_1+d{\hat\sigma}^{virtual},
\eeq
as in (\ref{rdf_ns2}).

The reason why we include the two diagrams $F_1$ and $F_2$ in this partial 
contribution is because it is only after adding these two contributions 
to the $G_i$ diagrams that the corresponding cross section factorizes. 
Notice that the two diagrams $G_4$ and $G_8$ generate collinear singularities 
which are different from those coming from the remaining diagrams 
of the same set. In fact, while in most of the G-diagrams the gluon 
can become collinear to an incoming quark line, in $G_4$ and $G_8$ we have 
a gluon splitting into two gluons which can become collinear. The latter 
singularity is characterized by the 
invariant $s_{23}\to 0$, and it appears also in the two 
diagrams denoted $F_1$ and $F_2$ from the $F$-set. 
These two singularities do not require factorization since they are directly 
canceled by corresponding ones in the virtual contribution. The virtual 
diagrams responsible of this 
cancelation involve the gluon self energy corrections to $L_1$ and $L_2$ in 
the final state, and the emissions of gluons, ghosts, $q\bar{q}$ and 
$q'\bar{q}'$ pairs in the final state, as shown in Fig.~7.  As noted by EMP, 
all the $explicit$ 
singularities in the real emission processes come from this mixed F-G set of 
diagrams. We use the word $explicit$ because there are other singularities due 
to collinear emission in the remaining sets $F_5,...,F_8$ and $H_1,...,H_8$ 
but they cancel out when the non-singlet combination is formed and need not 
be factorized. 

In a scheme with a totally anticommuting $\gamma_5$ we would clearly get a 
selection rule for the helicity contributions of the form, $M_{++}=0$. 
In fact all the final state emissions are vector-like and an incoming antiquark 
of positive helicity can be replaced by a quark of negative helicity 
flowing back toward the initial state. 

If it is postulated that $\gamma_5$ anticommute with the other 4-dimensional 
Dirac matrices and anticommute with the remaining ones, then
all the corresponding algebraic relations can be shown to be 
consistent with dimensional regularization \cite{bm}.
Therefore we  {\em define}
\beqa
&& \gamma_5\widehat{\widehat{\gamma}}_\mu + \widehat{\widehat{\gamma}}_\mu 
\gamma_5=0
\nonumber \\
&& \gamma_5\widehat{\gamma}_\mu - \widehat{\gamma}_\mu\gamma_5=0.
\nonumber \\
\eeqa
Application of this definition of $\gamma_5$ leads to the subdivision of 
all the kinematic momenta into 2 sets : 4 dimensional ones 
and $n-4$ dimensional ones, 
the latter called {\em hat-momenta}. All the dependence on the hat-momenta 
can be removed from the virtual corrections. The hat-momenta are 
removed only after performing the internal loop integrations 
using dimensional regularization. Notice that 
this procedure introduces spurious finite terms which  survive even in the 
limit in which $\epsilon\to 0$ $(n=4 - 2 \epsilon)$. These terms are due 
to the fact that single infrared poles in $1/\epsilon$ can be canceled by 
$\epsilon$-dependent factors generated by the trace in 
the $n-4$-dimensional subspace. There are other special features of this 
regularization which deserve special attention. 

In the $\overline{MS}$ 
scheme, the HVBM prescription for the regularization of the 
chiral states generates the  helicity non-conserving term 
shown in (\ref{rdf_sig1h}). In general one would expect that all the infrared 
sensitive contributions in the real emission diagrams, and therefore the 
G, the F and the H diagrams, would similarly contribute to an  
helicity non-conserving term. However, as we are going to show in the next 
section, this is not the case. 
The issue of the helicity conservation is completely settled by the $explicit$ 
singularities and by the diagrams of the first set ($d\hat\sigma_1$). 
This result suggests that 
the helicity non-conservation is not connected to all the singularities, but 
only to those singularities which require factorization. 
This observation is crucial in order to simplify the final result for 
both the virtual and the real contributions and identify their factorized 
form.  The factorized $d\hat\sigma_1$ cross section can be written in the form
\beq
s{d\hat\sigma_1^{fact}\over dtdu}(s,t,u,h)=s{d\hat\sigma_{1R}\over dtdu}(s,t,u,h)
+h\ s{d\hat\sigma^{hat}_0\over dtdu}(s,t,u),
\label{rdf_dfact1}
\eeq
and combining with the virtual contribution gives
\beq
s{d\tilde\sigma_1\over dtdu}(s,t,u,h)=s{d\tilde\sigma_{1R}\over dtdu}(s,t,u,h)
+h\ s{d\hat\sigma^{hat}_0\over dtdu}(s,t,u),
\label{rdf_dfact2}
\eeq
where
\beq
s{d\tilde\sigma_{1R}\over dtdu}(s,t,u,h)=s{d\hat\sigma_{1R}\over dtdu}(s,t,u,h)
+s{d\hat\sigma_1^{virtual}\over dtdu}(s,t,u,h).
\label{rdf_dfact3}
\eeq
In the HVBM prescription both $d\hat\sigma_{1R}$ and $d\hat\sigma_1^{virtual}$ are 
helicity non-conserving, but the sum  $d\tilde\sigma_{1R}$ conserves helicity.  
Namely,
\beq
s{d\tilde\sigma_{1R}\over dtdu}(s,t,u,h)=
(1-h)\ s{d\tilde\sigma_{1R}^\Sigma\over dtdu}(s,t,u),
\label{rdf_hcons}
\eeq
where $d\tilde\sigma_{1R}^\Sigma$  is the spin-averaged ({\it unpolarized}) 
result.
Notice that since $d\hat\sigma^{hat}$ is proportional to $h$ it does not 
contribute to the spin-averaged cross section which comes from setting
$h=0$.

We give the result for the non-singlet cross section in two different 
schemes: the $\overline{MS}$ scheme and the so called 
$\overline{MS}_p$ scheme \cite{GV}. In this later scheme we subtract 
additional $O(\epsilon)$ contributions to the polarized $\Delta P_{qq}$ 
splitting function, resulting in
\beq
\Delta P_{qq}(z)=P_{qq}(z)+4\epsilon(1-z).
\eeq
In this scheme helicity conservation at the parton level is maintained 
($d\hat\sigma^{hat}_0=0$) but $P_{qq}\neq\Delta P_{qq}$. In the 
$\overline{MS}$ scheme $P_{qq}=\Delta P_{qq}$ and helicity is not conserved 
at the parton level   ($d\hat\sigma^{hat}_0\
\neq 0$ ).
Since we move from one scheme to the other by a finite renormalization, the 
difference between the two evolutions is absorbed in the $O(\alpha_s)$ 
(i.e. $Z^{(1)}$) contribution to the renormalization factor in front of the 
parton distributions. These changes imply that the NLO contribution to the 
kernel (i.e. $P^{(1)}$) is also modified by an amount $P^{(1)}\to P^{(1)} 
- \beta_0 Z^{(1)}$, where $\beta_0$ is the first coefficient of the 
beta function (see sect. 10.2 of ref. \cite{FP} for more details).   

\subsection{Factorization of the Real Contributions and $d\hat\sigma_1$}

The standard procedure to be used in the calculation of the 
NLO contributions is to add the real emission to the virtual contributions, 
thereby canceling those singularities which are characterized by double pole 
in $1/\epsilon$. After factorization of the collinear singularities 
in the initial and final states the final result of the calculation 
is expressed in the form of regular terms and of plus distribution in the 
the variable $s_{2}$ in (\ref{rdf_sum1}). The two ``plus-distributions" in 
question are $1/(s_2)_{A+}$ and $(\ln(s_2)/s_2)_{A+}$, which are defined 
in terms of the upper limit of integration  in (\ref{rdf_eq2}) and defined in 
(\ref{rdf_defa}). 
These two distributions are defined by 
\beqa
&& \int_0^A d\,s_2 f(s_2){1\over (s_2)_{A+}}f (s_2)=\int_0^A {1\over s_2}
(f(s_2)- f(0)), \nonumber \\
&&\int_0^A ds_2\left( {\ln s_2\over s_2}\right)_{A+} f(s_2)= 
\int_0^A {\ln s_2\over s_2}\left(f(s_2)- f(0)\right).
\eeqa
As usual we define the relation between the bare and the renormalized 
structure functions by the equation 

\beq
G_{A\to i}(x,M^2)= \int_x^1 {d\,z\over z}\left[ \delta_{i j}\delta(z-1) +
{\alpha_s\over 2 \pi}R_{i\leftarrow j}(z,M^2)\right] G^{bare}_{A\to i}
\left({x\over z}\right).
\eeq
We choose for $R$ the usual form 
\beq
R_{i\leftarrow j}= -{1\over \epsilon}\Delta P_{i\leftarrow j}(z)
{\Gamma[1-\epsilon]\over \Gamma[1-2\epsilon]}\left({4 \pi \mu^2\over M^2}
\right)^{\epsilon} + C_{i\leftarrow j}(z)
\eeq
where the finite pieces $C_{i\leftarrow j}(z)$ are arbitrary. In the
$\overline{MS}$ 
scheme they are set to be zero. The polarized splitting functions, 
$\Delta P_{ij}$, are given by 
\beqa
&&\Delta P_{qq}=C_F\left[ {(1+z^2)\over (1-z)_+} + \frac{3}{2} \delta(1-z)
\right] \nonumber \\
&&\Delta P_{qg}=\frac{N_F}{2}[ 2 z-1]\nonumber\\
&&\Delta P_{gg}=N_C\left[ (1+z^4)\left(
\frac{1}{z}+\frac{1}{(1-z)_+}\right)-\frac{(1-z)^3}{z}
\right]+\frac{33-2 N_F}{6}\delta(1-z)\nonumber \\
&&\Delta P_{gq}=C_F[2-z],\nonumber 
\eeqa
with the distribution $1/(1-z)_+$ defined by 
\beq
\int_0^1 d\, z{ f(z)\over (1-z)_+}=\int_0^1{f(z)-f(1)\over 1-z}.
\eeq

The factorized cross section is given to order 
$\alpha_s^2$ by the renormalized cross section $d\hat\sigma_1$ with the 
the collinear initial and final state singularities subtracted
\begin{eqnarray}
s {d\hat\sigma_{ij}^{fact}\over d\,t d\,u} &=& 
s {d{\hat\sigma_1^{ij}}\over d\,t d\,u}\nonumber \\
 & - & {\alpha_s\over 2 \pi} \sum_k
\int_0^1 d\,z_1 R_{k\leftarrow i}(z_1,M^2)s {d\sigma^{(1)}_{k j}\over
d\, t}\mid_{p_1\to z_1 p_1}
\delta\left(z_1(s + t - Q^2) + u\right)\nonumber \\
&-& {\alpha_s\over 2 \pi} \sum_k
\int_0^1 d\,z_2 R_{k\leftarrow j}(z_2,M^2)s {d\sigma^{(1)}_{ik}\over d\,
t}\mid_{p_2\to z_2 p_2}
\delta(z_2(s + u - Q^2) + t).
\label{fact}
\end{eqnarray}

The collinear and soft divergences of the two unobserved jets are described 
by the limit $s_2\to 0$ and will appear dimensionally regulated in the form 
$s_2^{-1-\epsilon}$. These divergences can be expanded in terms of 
poles in $1/\epsilon$ and regulated distributions by the identity
\beq
{1\over s^{1+\epsilon}}=-{1\over \epsilon}\delta(s_2)\left[1 - \epsilon \ln A +
{1\over 2} \epsilon^2 \ln^2 A\right] + 
{1\over (s_2)_{A+}}-\epsilon \left({\ln s_2\over s_2}\right)_{A +} 
\eeq
up to order $\epsilon^2$. The collinear singularities associated with
the unobserved jets cancel when the contributions from the three quark-antiquark
annihilation processes $q\bar{q}\rightarrow gg \gamma^*$, 
$q\bar{q}\rightarrow q'\bar{q}' 
\gamma^*$ and $q\bar{q}\rightarrow q\bar{q} \gamma^*$ are added together. The 
soft component cancels against the corresponding poles from virtual 
contribution. Adding real and virtual contributions and factorizing the 
remaining collinear singularities with eq.~(\ref{fact}), all the divergences 
cancel and the result is free of all singularities. 

Details of the three particle phase space integration are given in
appendix A. In all cases the polarized and unpolarized three-body matrix
elements are integrated simultaneously since, except for the presence of
the hat-momenta present for the set of diagrams labeled $G$ and $F_1$
and $F_2$, they have a similar structure. 

The diagrams $|\sum_i G_i|^2$ and $|F_1+F_2|^2$ are integrated in
$4-2 \epsilon$ dimensions since they generate soft and collinear, and
collinear only poles respectively. The non-$4$-dimensional parts of these
matrix elements are different in the polarized and unpolarized cases in
the HVBM scheme. This leads to a different structure for the integrated
three-body matrix elements. In both cases there are
also finite contributions from the hat-momentum integrals.       
In the case of the diagrams $|F_1+F_2|^2$ the polarized and unpolarized
results differ by 
\begin{equation}
\delta \sigma \sim 4 N_f \delta(s_2)\frac{ Q^2 s+2 t^2+2 t u+2
u^2}{3 t u}
\end{equation}
where external color factors and coupling constants have been neglected.
This difference comes both from the $O(\epsilon)$ part of the three
body matrix element when multiplied by the factor
$-\delta(s_2)/\epsilon$ coming from the expansion of
$s_2^{-1-\epsilon}$ as well as from the hat-momenta. 

Similarly for the $|\sum_i G_i|^2$ terms there are differences between the
polarized and unpolarized result arising from the differences in the
matrix elements up to $O(\epsilon^2)$ as well as the hat-momentum
contributions. In this case the differences coming from the
$O(\epsilon)$ part of the matrix element are too complex to present
here.  The total contribution coming from the hat-momenta in the 
$\overline{MS}$ is given by
\beq
s{d\hat\sigma^{hat}\over dtdu}(s,t,u,h)=h\ s{d\hat\sigma_0^{hat}\over dtdu}(s,t,u),
\eeq
where
\begin{eqnarray}
\!\!&&\!\!\!\!\!\!\!\!\!s\frac{d\hat\sigma_0^{hat}}{dtdu}(s,t,u)
=-e^2_f K_2\frac{\alpha_s^2}{\pi s}\left\{
\frac{11}{8} N_C \delta(s_2)  \frac{(t + u)^2}{t u}
    -2 C_F s_2 \left[
\frac{u-3Q^2 }{u(Q^2 s_2 -t u)}
\right.
\right.
\nonumber\\ &&
+\frac{1}{(s_2-t)^2}
\left(\frac{ 2 s_2 t + Q^2 (s_2 + t)}{t^2} 
-  \frac{2Q^2}{u} \left(1 - \frac{2 s_2 t}{(s_2 - u)^2} \right) 
\right.
\nonumber\\ &&
\left.
\left.  \left.
+ \frac{1}{Q^2 s_2 - t u}\left( Q^4 + u (2 s_2 + u) 
- \frac{2Q^2s_2( 2  u s s_2 +  Q^2 (t-u)^2 )}{u (s_2 - u)^2}\right) \right)
+\{t\leftrightarrow u\}\right]
\right\}.
\end{eqnarray}

When the three-body contributions from $|F_1+F_2|^2$ and the $|\sum_i G_i|^2$ 
diagrams are all added together and the collinear poles
factorized according to (\ref{fact}) then the result is
given by

\begin{eqnarray}
\!\!&&\!\!\!\!\!\!\!\!\!s\frac{d\hat\sigma_{1R}}{dtdu}(s,t,u,h)
= e^2_f K_2\frac{\alpha_s}{s}
\left\{ \left((1-h)T_B-h\epsilon A_B\right)\delta(s_2) 
\left[ \frac{\alpha_s}{2\pi}\frac{\Gamma(1-\epsilon)}{\Gamma
(1- 2 \epsilon)}\left(\frac{4\pi \mu^2}{Q^2}\right)^{\epsilon}\right.\right.
\nonumber \\
 &  & \times\left(\frac{2 C_F + N_C}{\epsilon^2} 
+ \frac{1}{\epsilon}
\left(3 C_F - 2 C_F \ln \frac{s}{Q^2} + \frac{11}{6} N_C + N_C \ln
\frac{sQ^2}{ut} -\frac{1}{3}N_F\right)\right)
\nonumber \\ &&
+\frac{\alpha_s}{2\pi}\left\{
       \left(\frac{11}{6} N_C - \frac{1}{3} N_F\right) 
\ln\left(\frac{Q^2 }{A}\right) + \left( 
\frac{67}{18} N_C - \frac{5}{9} N_F\right) + N_C \ln\left(\frac{Q^2 }{A}\right)^2  
\right.
\nonumber \\ &&
\left.\left.
   +\left(C_F- \frac{1}{2}N_C\right)\left( \frac{\pi^2}{3} 
+ \ln\left(\frac{A^2 s}{Q^2 u t}\right)^2 \right) 
\right\}\right]
+\frac{\alpha_s}{2\pi}(1-h)T_B \left[ \frac{ 2N_F - 11 N_C}
{6 (s_2)_{A+}} 
\right.
\nonumber \\ &&
  + (2C_F-  N_C)\left(
2\left(\frac{\ln\left(s_2/A\right)}{s_2}\right)_{A+} + \frac{1}{(s_2)_{A+}}
\ln\left(\frac{s^2 A^2}{(u-s_2)(t-s_2)(u t - Q^2 s_2)}\right) \right)
\nonumber \\ &&
\left.\left.
+  \frac{2C_F}{(s_2)_{A+}} \ln\left(\frac{u t - Q^2 s_2}{(u-s_2)(t-s_2)}\right) 
+ 4 C_F \left(\frac{\ln\left(s_2/Q^2\right)}{s_2}\right)_{A+} 
\right] \right\}
\nonumber \\ &&
+ e^2_f K_2\frac{\alpha_s^2}{s\pi} (1-h)
\left\{ \left[
 N_C \left( - \frac{11}{6} \frac{ s+Q^2}{t u} + \frac{s^2 ( 3s_2 - 4t)}{2 t u
( t-s_2)^2} - \frac{2 s}{u(t-s_2)} +
           \frac{Q^2}{3 t^2} \right) 
\right.\right.
\nonumber \\ &&
+C_F\left( \frac{s}{(t-s_2)^2} - \frac{2}{t-s_2} - \frac{s}{u t} 
+ \frac{2 Q^2 u + t s_2 -2 Q^2 s_2}{t(t u - Q^2 s_2)}
           +\frac{2 Q^2 (t -u)}{t u (t-s_2)} \right) 
\nonumber \\ &&
+ C_F \ln\left(\frac{Q^2}{s_2}\right)
\left( \frac{4 u t(u-Q^2) -4 Q^2(u-Q^2)^2 - u t s_2}{u t (u t- Q^2 s_2)} 
+ \frac{Q^2-u}{(t-s_2)^2}
\right.
\nonumber \\ &&
\left.\left.
+ \frac{2Q^2 -u}{t(t-s_2)} - \frac{2}{t}+ \frac{Q^2}{t^2}\right) 
+ (u\leftrightarrow t) \right]
\nonumber \\ &&
+ N_F \frac{2 s t u -Q^2 (t - u)^2 }{3 t^2 u^2}
+ 2(2 C_F - N_C)\ln
\left(\frac{s s_2}{(u-s_2)(t-s_2)}\right)
\frac{Q^2 + s}{ t u} 
\nonumber \\ &&
+ (2 C_F - N_C) \ln\left(\frac{s s_2}{(u t - Q^2 s_2)}\right) 
 \frac{(Q^2 -s) ((t-Q^2 )^2 + (u-Q^2 )^2)}{(t-s_2) t u (u-s_2 )}
\nonumber \\ &&
\left.+ 2 C_F \ln\left(\frac{u t - Q^2 s_2}{(u-s_2)(t-s_2)}\right) 
\frac{Q^4 + s^2 + (Q^2 - t)^2 + (Q^2 - u)^2}{(t u-Q^2 s_2) t u } \right\},
\end{eqnarray}
where 
\beq
s{d\hat\sigma_1^{fact}\over dtdu}(s,t,u,h)=s{d\hat\sigma_{1R}\over dtdu}
(s,t,u,h)+\ s{d\hat\sigma^{hat}_0\over dtdu}(s,t,u,h).
\eeq

Notice that the term that survives when $h=0$ in $d\hat\sigma_{1R}$ exactly
cancels the corresponding term in $d\hat\sigma{virtual}$, resulting in the 
helicity conservation implied by (\ref{rdf_hcons}).

\subsection{The Finite Cross Sections $d\hat\sigma_2$,  $d\hat\sigma_3$, 
and $d\hat\sigma_4$}

The remaining real contributions to the non-singlet cross section,
\begin{eqnarray}
d{\hat\sigma}_2&\sim& |F_3+F_4|^2,\\
d{\hat\sigma}_3&\sim& 2Re\left(\sum_{i=1}^4F_i\sum_{i=5}^8F^*_i\right), \\
d{\hat\sigma}_F&\sim& |\sum_{i=5}^8F_i|^2,
\end{eqnarray}
have neither collinear or soft singularities and
therefore may be integrated in $4$-dimensions. In this case the
helicity structure is trivially equal to that expected by helicity conservation.
These differential cross sections are as follows:

\begin{eqnarray}
\!\!&&\!\!\!\!\!\!\!\!s\frac{d\hat\sigma_2}{dtdu}(s,t,u,h) = e^2_f 
\frac{K_2}{s}\frac{\alpha_s^2}{\pi}
(1-h) C_F
\left\{ \frac{4}{s\lambda^4}\left[ (t+u)^2 ( (5  Q^2 - s) s_2 -t^2-u^2 )
\right.\right.
\nonumber \\
 &  &\left.\phantom{\frac{1}{1}}\hspace{-4mm}- 2 s_2 ( (5 Q^2 +s ) s_2 Q^2  + (Q^2  -3 s) t u )\right]
+ \frac{2}{s \lambda^5}
\ln\left(\frac{\rho +\lambda}{\rho -\lambda}\right)
\left[ \rho\lambda^2  (t^2+u^2)
\right.
\nonumber \\
&&
- 2 s_2  (12 (Q^2+s_2) s t u +  s(t + u)^3 -2 s_2  Q^4\rho )  
\nonumber \\
&&
+ 4 \frac{s_2}{\rho} \left\{ 
 2 s_2 \left( 3 Q^4 s s_2 + 3 s (s - s_2) t u  
     +2 Q^4 (2 Q^2 s + t u) - Q^2 s (Q^2 (t + u)+ t^2 + u^2) \right)
\right.  
\nonumber \\
&&
\left.\phantom{\frac{1}{1}} \left.  \left.  
-(Q^2+s) (t+u)^2 t u
\right\}\right]\right\},
\end{eqnarray}
where 
\begin{equation}
\lambda=\sqrt{(t+u)^2 - 4 Q^2 s_2},\ \ \ \rho=Q^2+s-s_2.
\end{equation}

\begin{eqnarray}
\!\!&&\!\!\!\!\!\!\!\!s\frac{d\hat\sigma_3}{dtdu}(s,t,u,h) = e^2_f 
\frac{K_2}{s}\frac{\alpha_s^2}{2\pi}
(1-h) (2C_F -N_C)
\left\{ \left[ \frac{3 us - 6 Q^2 s - 8 s_2 Q^2}{2 s u^2}
\right.\right.
\nonumber \\
&&\left.
+\frac{t-u-s}{2 s_2 u} + \frac{s^2 }{t(u-s_2)^2}+\frac{5t-3u}{2 su}
-\frac{3s^2+2 tu}{tu(u-s_2)}\right]
\nonumber \\
& &+ \frac{1}{ \lambda^2}\left[
\frac{2 t s_2 (t-u-s_2)}{s u} +\frac{2 st^2}{us_2}+ \frac{2s_2^2}{s}
+\frac{t(2t-3s-s_2)}{u}+(s-s_2)
\right.
\nonumber \\
&&\left.+\frac{t(s+s_2)(u^2-t^2)(\rho+3s)}{2 u s s_2\rho}\right]
\nonumber \\
& &+ \frac{3}{ \lambda^4}\frac{st^2(s+s_2)(t^2+tu-2u^2)}{2 us_2}\left[
\frac{1}{\lambda}\ln\left(\frac{\rho+\lambda}{\rho-\lambda}\right)
-\frac{2}{\rho}\right]
\nonumber \\
&&
+\frac{1}{\lambda}\ln\left(\frac{\rho+\lambda}{\rho-\lambda}\right)
\left[\frac{2t(t+s) +9 s^2}{2 us_2} +\frac{4 s_2^2 -2 s_2 t +t^2}{s u}
+\frac{5u-6s_2}{s}
\right.
\nonumber \\
&&+ \frac{3s+u}{s_2}+\frac{s-2Q^2}{2 u} +\frac{4s+2s_2}{\rho}
\nonumber \\
&& \left.
+ \frac{t}{\lambda^2}\left(t-Q^2-\frac{su}{s_2} +\frac{t(t-s_2-s)}{u}
+\frac{st(t-2s)}{us_2}\right)\right]
\nonumber \\
&& + 2\ln\left(\frac{su}{ s_2 t + Q^2(u -2 s_2)}\right)\frac{s_2^2+(u-s_2)^2}
{su\rho}
\nonumber \\
&&
\left.
+\ln\left(\frac{su^2}{Q^2(u-s_2)^2}\right)\left(\frac{2s+t+u}{su}
+\frac{2s^2+2st+t^2}{tus_2}-\frac{2 Q^2}{u^2}\right)\right\}
\nonumber \\
&&
+e_f^2 \frac{K_2}{s} \frac{\alpha_2^2}{2\pi}(2C_F-N_C)\{(u\leftrightarrow t)\}.
\end{eqnarray}

\begin{eqnarray}
\!\!&&\!\!\!\!\!\!\!\!s\frac{d\hat\sigma_4}{dtdu}(s,t,u,h) = e^2_f 
\frac{K_2}{s}\frac{\alpha_s^2}{\pi}
(1+h) (2C_F -N_C)
\left\{\frac{1}{stu}
\left[ 2 Q^2 s_2 \left(\frac{t}{u} + \frac{u}{t}\right)
\phantom{\frac{Q^2}{t}}\right.
\right.
\nonumber\\ && \left.-(t + u)^2 
+ \left(2 (Q^2 -s ) s_2 -  (t + u)^2\right)
\ln\left(\frac{Q^2 s_2}{t u}\right) \right] 
\nonumber \\
&&+ \left[\left\{
\left(\frac{2 s^2}{\rho t(t-s_2) }
+ \frac{ Q^2-s-s_2}{t(t-s_2)}\right) 
   \ln\left(\frac{(t-s_2)\rho - s t}{st}\right) 
\right.
\right. \nonumber\\
&&
\left.\left.
\left.
+\frac{ Q^2}{t^2} \ln\left(\frac{s t^2}{Q^2 (s_2 - t)^2}\right) \right\}
+ \{u \leftrightarrow t\} \right]\right\}.
\end{eqnarray}

\section{Conclusions}

We have presented the parton-level analytical results for 
the next-to-leading order non-singlet virtual corrections to the 
Drell-Yan differential cross-section.  
The dependence of the differential cross section on the helicity of the 
initial state partons is shown explicitly (the spins of the final state 
partons are summed). Although the calculation is very involved 
due to the presence of chiral projectors in the initial state, the result 
is quite compact and has been presented in a form from which 
cancelation of the infrared, collinear and infrared plus collinear 
singularities is evident. The non-singlet parton level asymmetry defined by,
\beq
s{d\hat\sigma_{LL}\over dt}={1\over2}\left(s{d\hat\sigma_{++}\over 
dt}-s{d\hat\sigma_{+-}\over dt}\right),
\eeq
is given by order $\alpha_s^2$ by,
\beq
s{d{\hat\sigma}^{LL}_{NS}\over dtdu}(s,t,u)=
-s{d{\hat\sigma}^\Sigma_{NS}\over dtdu}+
s{d{\hat\sigma}^{hat}_0\over dtdu}-
2s{d{\hat\sigma}^\Sigma_{4}\over dtdu},
\eeq
where $d{\hat\sigma}^\Sigma_{NS}$ is the non-singlet {\it unpolarized}
spin-averaged cross section given by
\beq
s{d{\hat\sigma}^\Sigma_{NS}\over dtdu}(s,t,u)=
s{d{\hat\sigma}_{B}^\Sigma\over dtdu}+
s{d{\tilde\sigma}_{1}^\Sigma\over dtdu}+
s{d{\hat\sigma}_{2}^\Sigma\over dtdu}+
s{d{\hat\sigma}_{3}^\Sigma\over dtdu}-
s{d{\hat\sigma}_{4}^\Sigma\over dtdu},
\eeq
where $d\hat\sigma_B$ is the order $\alpha_s$ Born term and the remaining 
cross sections are of
order $\alpha_s^2$.
The leading order parton level result that
\beq
\hat A_{LL}={d\hat\sigma_{LL}\over dt}\bigg/{d\hat\sigma^\Sigma\over dt}=-1.
\eeq
is not true at order $\alpha_s^2$ in the $\overline{MS}$ scheme.  In this
scheme, $P_{qq}(z)=\Delta P_{qq}(z)$ in D-dimensions and the helicity 
non-conserving
term, $d\hat\sigma^{hat}_0$ is {\it explicitly} needed in the convolution of the
hard scattering with the NLO evolved structure functions.  The evolution DGLAP kernel 
has obviously to be evaluated in this same scheme.

Our unpolarized cross section, gotten by setting $h=0$, agrees with the previous
result of Ellis, Martinelli and Petronzio \cite{EMP} in the non-singlet sector.
Our calculation is a first step toward the extension of the classical 
$O(\alpha_s^2)$ Drell-Yan result to the case of longitudinal polarized
beams. A complementary discussion of our results and of the methods 
developed by us in the analysis of the virtual corrections 
will be presented elsewhere \cite{CCG}.

\section{Acknowledgments}
We thank R.K. Ellis, G. Bodwin and G. Ramsey for discussions.
C.C. thanks Alan White and the Theory Group at Argonne, 
the Theory group at Jefferson Lab. for hospitality
and R. Gantz for encouragement. 

\section*{Appendix A: 2- and 3-particle Phase Space for Polarized Drell Yan} 

Let's start considering the virtual contributions to the cross section.
The relevant phase space integral is given by
\begin{equation}
PS_2=\int \frac{d^nqd^nk_2}{(2\pi)^{n-1}(2\pi)^{n-1}}\delta_+(q^2-Q^2)
\delta_+(k_2^2)\delta^n(p_1+p_2-k_2-q) (2\pi)^n.
\end{equation}
It is convenient to introduce light-cone variables
$(q^+,q^-,q_\perp)$, with
\begin{equation}
q=q^+n^+ + q^-n^- + q_\perp
\end{equation}
and
\begin{equation}
n^\pm\equiv \frac{1}{\sqrt{2}}(1, 0_\perp, \pm 1)
\end{equation}
and work in the frame
\begin{equation}
p_1^+=p_2^-=\sqrt{\frac{s}{2}}
\end{equation}
to get
\begin{eqnarray}
PS_2 &=  & \int \frac{d^+qd^-q}{(2\pi)^{n-2}}|q_\perp^2|^{n/2-2} d|q_\perp^2|
\delta(2q^+q^- -|q_\perp^2|-Q^2)\delta(s+t+u-Q^2)\Omega^{n-3}\\
 & = &  \int d\,q^+d\,q^-(2q^+q^-- Q^2)^{n/2-2}\delta(s+t+u-Q^2)
\frac{\Omega^{n-3}}{(2\pi)^{n-2}}
\end{eqnarray}
with 
\begin{equation}
\Omega^{n-3}\equiv 2\int^\pi_0 \prod^{n-3}_{l=1} \sin\theta_l^{n-l-3} d\theta_l
=\frac{2 \pi^{n/2-1}}{\Gamma(n/2-1)}
\end{equation}
using
\begin{equation}
\frac{\partial(q^+,q^-)}{\partial(t,u)}=\frac{1}{2s}
\end{equation}
and after covariantization
\begin{equation}
(2q^+q^- -Q^2)\rightarrow \frac{(Q^2-t)(Q^2-u) -s Q^2}{s} =\frac{ut}{s}
\end{equation}
\begin{equation}
PS_2 =\int dt du \frac{1}{2s}\left(\frac{ut}{s}\right)^{n/2-2}
\frac{2 \pi^{n/2-1}}{\Gamma(n/2-1)}
\frac{\delta(s+t+u-Q^2)}{(2\pi)^{n-2}}
\end{equation}
\begin{equation}
\sigma= \frac{1}{4 N_c}\int \frac{dtdu}{(2s)^2} 4\pi \alpha_s 
\left(\frac{\mu^2}{Q^2}\right)^\epsilon
\left(\frac{sQ^2}{ut}\right)^\epsilon \delta(s+t+u-Q^2)T_0(Q^2,u,t) .
\end{equation}
Moving to the contributions with 3 particles in the final state, we introduce 
the invariants

\beqa
&& s=(p_1 + p_2)^2\nonumber \\
&& s_{12}=(k_1 + k_2)^2 \,\,\,\, s_{23}=(k_2 +k_3)^2\nonumber \\
&& t_1=(p_1-k_1)^2 \,\,\, u_1=(p_2-k_1)^2 \nonumber \\
&& s_{13}=(k_1 + k_3)^2 \nonumber \\
&& t_2=(p_1-k_2)^2 \,\,\,t_3=(p_1-k_3)^2 \nonumber \\
&&  u_2=(p_2-k_2)^2 \,\,\,\,\, u_3=(p_2-k_3)^2. 
\eeqa
Only five of them are independent since 
\beqa
&& s=Q^2 -(u_1 +u_2 +u_3)= Q^2-(t_1 + t_2 +t_3 )=s_{12} +s_{23} + 
s_{13} - Q^2
\nonumber \\
&& s_{12}=s + u_3 + t_3 \nonumber \\
&& s_{23}=s + u_1 + t_1 -  Q^2 \nonumber \\
&& s_{13}=s + u_2 + t_2. 
\eeqa

A discussion of the modifications which appear in the integration over the 
phase space of the 3 final states, due to this ansatz, is presented in the 
next section.

The use of the t'Hooft-Veltman regularization 
\cite{thv} introduces a dependence of the matrix elements on the hat-momenta 
which requires, in part, a modification of the phase space integral which 
appear in the unpolarized case. 

In the case of unpolarized scattering,
 the singularities are generated by poles 
in the matrix elements which have the form $1/t_3$, $1/u_3$, $1/(t_3 u_3)$ and 
similar ones, in multiple combinations of them. Multiple poles can be reduced 
to sums of combinations of double poles by using simple identities  
among all the invariants and by the repeated use of partial fractioning. 
This is by now a well established procedure. 
In our case we encounter new terms of the 
form $1/t_3^2$ and $1/u_3^2$ and new matrix elements containing typical factors 
of the form ${\widehat{\widehat{k}}}_3$,  
$\widehat{\widehat{k}}_2$, and $\widehat{\widehat{k_2\cdot k_3}}$
at the numerator. Let's discuss for a moment these last terms containing 
hat-momenta. It is obvious that by a suitable choice of the parameterizations 
given by the sets 1, 2 ,3 and 4, (defined in the next section) we are able 
to reduce to the ordinary phase 
space result given by (\ref{ps3}) all the matrix elements containing scalar 
products of the form $\widehat{\widehat{p_i\cdot k_j}}$, 
$\widehat{\widehat{k_1\cdot k_j}}$ 
with $i=1,2$, $j\,=1,\,2,\,3$. Therefore it is possible to set to zero, 
after taking the traces, all the products that contain such combinations of
 hat-momenta. Then, the only 
matrix elements of hat-momenta which are left and which are not 
set to zero are those containing products of the form 
$\widehat{\widehat{k_a\cdot k_b}}$,with $a,\,b=\,1,\,2$. 

The generic 3-particle phase space integral is given by the formula
\beqa
PS_3&\equiv& 
f^{stat} \int {d^n k_1\over (2 \pi)^{n-1}} 
{d^n k_3\over (2 \pi)^{n-1}}{ d^n k_2\over (2 \pi)^{n-1}} \nonumber \\
& & \delta(k_3^2)\delta(k_2^2)
\delta(k_1^2-Q^2) (2 \pi)^n\delta^n(p_1 +p_2 - k_3 -k_2 -k_1).
\label{ps1}
\eeqa
Care must be taken in integrating over $k_2$ and $k_3$ when they are identical 
partons.  In this case, an extra factor of $1/2$ must be inserted so as not to 
double count the number of identical partons.  We have indicated this by 
including a statistical factor $f^{stat}$ that is equal to $1/2$ for identical 
partons, otherwise it is equal to one.  We use the notation $k_1\equiv q$ to 
characterize the momentum of the 
virtual photon and we lump together the momenta $k_3$ and $k_2$ as follows

\beqa
&& PS_3={f^{stat}\over (2 \pi)^{2 n -3}}
\int d^n k_1 d^n k_3 d^n k_2 d^n k_{23} \delta(k_3^2)\delta(k_2^2)
\delta(k_1^2- Q^2)\delta^n(p_1 +p_2  -k_2- k_1)\nonumber \\
&& \delta( p_1 + p_2 -k_1 - k_{23})
\delta^n(k_3 + k_2 - k_{23})\nonumber \\
&&= {f^{stat}\over (2 \pi)^{2 n -3}}\int d^n q d^n k_{23}\delta(q^2 - Q^2)
\delta^n(p_1 + p_2 - q - k_{23})\times PS_2,
\label{ps2}
\eeqa
where
\beqa
&& PS_2\equiv \int d^n k_2\delta(k_2^2)\delta((k_{23}- k_2)^2)
\nonumber \\
&& ={(k_{23}^2)^{n/2-2}\over 2^{n-2}}
{\pi^{n/2-3/2}\over \Gamma[n/2-3/2]}\int_{0}^{\pi}d\theta_1 
\sin^{n-3}\theta_1\int_{0}^{\pi}d\theta_2 \sin^{n-4}\theta_2
\label{trickp2}.
\eeqa
We obtain 
\beqa
PS_3={f^{stat}\over (2 \pi)^{2 n -3}}{\pi^{n/2-3/2}\over 2^{n-2}
\Gamma[n/2-3/2]}\int d^n q\delta(q^2-Q^2)(p_1 + p_2-q)^2]^{n/2-2}
\times I_{1,2} 
\eeqa
where 
\beq
I_{1,2}=\int_{0}^{\pi}d\theta_1 \sin^{n-3}\theta_1
\int_{0}^{\pi}d\theta_2 \sin^{n-4}\theta_2.
\eeq
We now use the light-cone parameterization of $q$ exactly as in the 
case of the derivation of the 2-to-2 cross section, perform the $q_\perp$ 
integration and switch to $t,u$ variables to finally get 
\beq
PS_3={(4\pi)^{2 \epsilon}f^{stat}\over 2^8 \pi^4 s \Gamma[1-2 \epsilon]}
\left({s\over s_{23}(u t - Q^2 s_{23})}\right)^{\epsilon} I_{1,2},
\label{ps3}
\eeq
where we have used the relation $s+t + u= s_{23} + Q^2$ and the covariantization
\beq
2 q^+ q^- - Q^2\to {u t - Q^2 s_{23}\over s}.
\eeq

\subsection*{A.1: Integration of the Real Emission Diagrams}

In order to integrate over the matrix elements, we need to evaluate the 
various scalar products which appear in such matrix elements,  in the 
c.m. frame of the pair $(1,2)$. For this purpose we define the 
functions 

\beqa
&&\lambda(x,y,z)= x^2 + y^2 +z^2-2 x y -2 y z -2 x z \nonumber \\
&& P[x,y,z]={\lambda^{1/2}(x,y,z)\over 2 \sqrt{x}} \nonumber \\
&& E[x,y,z] ={x+y-z\over 2 \sqrt{x}}.
\eeqa
It is easy to show that 

\beqa
&& |\vec{p}_1|=P[s_{23}, p_1^2,u_1]={s_{23}- u_1\over 2 \sqrt{s_{23}}}
\nonumber \\
&& |\vec{p}_2|=P[s_{23},p_2^2,t_1]=
{s_{23}- t_1\over 2 \sqrt{s_{23}}} \nonumber \\
&&|\vec{k}_3|=|\vec{k}_2|=P[s_{23},p_1^2,p_2^2]=
{\sqrt{s_{23}}\over 2}\nonumber \\
&& |\vec{k}_1|=\sqrt{s\over s_{23}}P[s,Q^2,s_{23}]=
{\sqrt{\lambda(s, Q^2,s_{23})}\over 2 \sqrt{s_{23}}}\nonumber \\
&& p_1^0=E[s_{23},p_1^2,u_1]={s_{23}- u_1\over 2 \sqrt{s_{23}}}\nonumber \\
&& p_2^0=E[s_{23},p_2^2,t_1]={s_{23}- t_1\over 2 \sqrt{s_{23}}}.
\label{list}
\eeqa

We get
\beqa
&& p_1^0={s_{23}-u_1\over 2 \sqrt{s_{23}}}\nonumber \\
&& p_2^0={s_{23}-t_1\over 2 \sqrt{s_{23}}}\nonumber \\
&& k_1^0={Q^2 -s + s_{23}\over 2 \sqrt{s_{23}}}\nonumber \\
&& k_3^0=k_2^0={\sqrt{s_{23}}\over 2}.
\label{k3}
\eeqa

In the derivation of (\ref{k3}) we have used the relation 
$s+ t_1 + u_1= Q^2 + s_{23}$. 
There are four different parameterizations of the integration momenta 
which we will be using. In the first one, which is suitable for unpolarized 
scattering one defines (in the c.m. frame of the pair $(1,2)$)
in general 

\beqa
&& k_3={\sqrt{s_{23}}\over 2}(1,...,\cos\theta_2 \sin\theta_1,\cos\theta_1)
\nonumber \\
&& k_2={\sqrt{s_{23}}\over 2}(1,...,-\cos\theta_2 \sin\theta_1,-\cos\theta_1)
\nonumber \\
&& p_1=p_1^0(1,0,...,0,\sin\psi_1,\cos\psi_1)\nonumber \\
&& p_2=p_2^0(1,0,...0,\sin\psi_1,\cos\psi_1)\nonumber \\
&& k_1=(k_1^0,0,...0,|k_1|\sin \psi_2, |k_1|\cos \psi_2),
\label{kk}
\eeqa
where the dots denote the remaining $n-2$ polar components. 

It is convenient to use the parameterizations 
\begin{itemize}
\item{set 1}
\beqa
&& p_1=p_1^0(1,0,...,0,0,1)\nonumber \\
&& p_2=p_2^0(1,0,...,-\sin\psi'',0,\cos\psi'')\nonumber \\
&& k_1=(k_1^0,0,...,-|\vec{k}_1|\sin\psi,0,|\vec{k}_1|\cos\psi),
\eeqa

\item{set 2}
\beqa
&& p_1=p_1^0(1,0,...,\sin\psi'',0,\cos\psi'')\nonumber \\
&& p_2=p_2^0(1,0,...,0,0,1)\nonumber \\
&& k_1=(k_1^0,0,...,|\vec{k}_1|\sin\psi',0,|\vec{k}_1|\cos\psi'),
\eeqa

\item{set 3}
\beqa
&& p_1=p_1^0(1,0,...,\sin\psi,0,\cos\psi)\nonumber \\
&& p_2=p_2^0((1,0,...,-\sin\psi',0,\cos\psi')\nonumber \\
&& k_1=(k_1^0,0,...,0,0,|\vec{k}_1|),
\eeqa
\end{itemize}
where $0,...$ refers to $n-5$ components identically zero. 
It is straightforward to obtain the relations 

\beqa
&& \cos\psi''={(s_{23}-t_1)(s_{23}- u_1) - 2 s_{23} s\over (s_{23}- t_1)(s_{23}- u_1)}\nonumber \\
&& \cos\psi={(Q^2 - s + s_{23})(s_{23}- u_1)-2 s_{23}(Q^2 - t_1)\over 
\lambda^{1/2}(s,Q^2,s_{23})(s_{23}- u_1)}
\nonumber \\
&& \cos\psi'={(s_{23}- t_1)(Q^2 - s + s_{23}) -2 s_{23}(Q^2 - t_1)\over 
\lambda^{1/2}(s,Q^2,s_{23})}.
\eeqa

We select a specific set depending upon the 
form of the hat-momenta in the matrix elements.
We choose the center of mass of the gluon pair 
$(2,3)$. 
Following the notation of references \cite{thomas,bennaker} we define 

\beqa
&& I_n^{(k,l)}\equiv \int_0^\pi d\,\theta_1\,\int_0^\pi\, d\,\theta_2 
\, \sin\,\theta_2^{n-4}\left(a + b \cos\,\theta_1\right)^{-k}\nonumber \\
&& \times \left(A + B \cos\,\theta_1 + C \sin\,\theta_1 \cos\,\theta_2
\right)^{-l}, 
\label{ben}
\eeqa
where a,b, A, B, C, are functions of the external kinematic variables . 

This expression trivially contains collinear singularities if $k\geq 1,\,\, 
l\geq 1$ and $a^2=b^2$ and/or $A^2 = B^2 + C^2$. The singularities are traced 
back to the emission of massless gluons from the initial state and in the final 
states. In this special case, it is convenient to rescale the integral 
and let the angular variables $\psi,\, \psi',\, \psi''$ defined above appear. 
 The cases 
$a^2=b^2$ and $A^2\neq B^2 + C^2$, $a^2\neq b^2$ and $A^2 = B^2 + C^2$
and $a^2\neq b^2$ and $A^2 \neq B^2 + C^2$ are discussed in 
ref.~\cite{bennaker}. It is easy to figure out from the structure of the 
matrix elements which cases require a four dimensional integration and which, 
instead, have to be evaluated in $n$ dimensions. This procedure is standard 
lore. Notice that only two independent angular variables 
$\theta_1$ and $\theta_2$ appear at the time.

Matrix elements containing more than 2 integration invariants 
(for instance $1/t_2 u_2 u_3$ or $1/t_2 t_3 u_2$) have to be partial 
fractioned by using the Mandelstam relations. 
Just to quote an example, 
we can reduce these ratios in a form suitable for integration 
in the $\theta_1$ $\theta_2$ variables by partial fractioning

\beq
{1\over t_2 u_2 u_3}= {1\over t_2 u_2 u_3}\left( {u_2 + u_3\over Q^2 - s - u_1}
\right)
\eeq
where we have used the identity 
\beq
{u_2 + u_3\over Q^2 -s - u_1}=1.
\eeq
If we let $I[.]$ denote the corresponding angular integral, we get 
\beq
I[1/t_2 u_2 u_3]= {1\over Q^2 -s - u_1}
\left(I[1/t_2 u_2] +I[1/t_2 u_3]\right).
\eeq
The integrals on the rhs of this equation are now in the standard 
$\theta_1$ $\theta_2$ form. 

Defining 

\beqa
&& V_n[\mp,\mp]\equiv \int_{0}^{\pi}\int_{0}^{\pi}d\theta_1 
d\theta_2 {\sin^{n-3}\theta_1\sin^{n-4}\theta_2\over 
(1- \cos\theta_1)^i(1\mp \cos\chi \cos\theta_1 \mp \sin\chi\cos\theta_2\sin 
\theta_1)^j}
\eeqa
we get
\beqa
&& V_n[-,-]= C[n,i,j]F[1,j,n/2-1,\cos^2{\chi\over 2}]\nonumber \\
&& V_n[+,-]=C[n,i,j]F[1,j,n/2-1,1/2 +\sin{\chi\over 2}]\nonumber \\
&& V_n[-,+]= V_n[-,-]\nonumber \\
&& V_n[+,+]= C[n,i,j]F[1,j,n/2-1,\sin^2{\chi\over 2}],
\eeqa

where
\beqa
C[n,i,j]\equiv 2^{1-i -j}\pi{\Gamma[n/2-1-j]\Gamma[n/2-1-i]\over 
\Gamma[n-2-i-j]}
{\Gamma[n-3]\over \Gamma^2[n/2-1]}.
\eeqa

\subsection*{A.2: Integration Over Hat-momenta}

Let's now consider the modifications induced by the HVBM prescription in the 
calculation of the real emission diagrams.

\beq
PS_2\equiv \int d^n k_2\delta(k_2^2)\delta((k_{23}- k_2)^2)
\widehat{\widehat{k}}_2^2
\label{trickp3}.
\eeq
We have set $k_2=(\widehat{\widehat{k}}_2,\widehat{k}_2)$, with
\beq
\widehat{\widehat{k}}_2=k_2^0(1,\cos\theta_3 \sin\theta_2 \sin\theta_1,
\cos\theta_2 \sin\,\theta_1,\cos\,\theta_1)
\eeq

being the 4-dimensional part of $k_2$. 
We easily get 
\beq
\widehat{\widehat{k}}^2_2= {s_{23}\over 4}
\sin^2\theta_3\,\sin^2\theta_1\,\sin^2\theta_2.
\eeq

Therefore the usual angular integration measure 
\beq
d\Omega^{(n-2)}=\prod_{l=1}^{n-2}\,\sin^{n-l-2}\theta_l\,d\theta_l 
\eeq
 
 is effectively modified to 
\beq
d\Omega^{(n-2)}=\prod_{l=1}^{3}\sin^{n-l}\theta_1 d\theta_l\times 
\prod^{n-2}_{l=4} \sin^{n-l-2}\theta_l d\theta_l. 
\eeq
The intermediate steps of the evaluation are similar to those in the previous 
section. An analogous treatment can be found in \cite{CG}, 
to which we refer for more details. 

In $n=4-2 \epsilon$ dimensions we get
\beq
PS_3={ \pi^{2\epsilon} \epsilon\over 2^8 \pi^4 \Gamma[1-\epsilon]}
\left( {u t - Q^2 s_{23}\over s}\right)^{-\epsilon}
{s_{23}^{1-\epsilon}\over 2} \int_0^{\pi}d\theta_1 \,
\sin\theta_1^{3- 2\epsilon} \,\int_0^\pi \,d\theta_2 \, 
sin\theta_2^{2-2\epsilon}.
\eeq

Therefore, in the HVBM scheme, additional contributions are generated by the 
integration over the hat-momenta in the 3 parton phase space,  which are not 
present in other regularization schemes. These contributions are either 
finite or of order $1/\epsilon$ and therefore modify the subtraction of the 
collinear terms in the total (real + virtual) cross section.  

\section*{Appendix B: Some matrix elements}

The list of the integrals relevant for the real emission diagrams can be 
found in ref~ \cite{bennaker}. We have independently 
checked all those integrals which are needed for our calculation 
and we have found agreement with the authors. Except for the integrals 
requiring an n-dimensional integration measure $(I_n^{(k,l)})$, the remaining 
integrals are quite straightforward. 

In our calculation we also encounter matrix elements of the 
form 
$ t_2/s_{12}$, $t_2/s_{12}^2$ and $t_2^2/s_{12}^2$. 

Using the reference frames discussed in this paper it is possible 
to reduce the angular integration of these matrix elements to integrals of 
the form 

\beq
Y_n^{(k,l)}=\int_0^{\pi} \sin\theta_1^{n-4}\int_0^\pi \sin\theta_2^{n-3}{ 
(1-\cos\theta_1)^k
\over(A + B \cos\theta_1 + C \sin\theta_1 \cos\theta_2)^l}.
\eeq

We can set $n=4$ since - due to the fact 
that the photon is massive - the integration is finite. 
For instance we get
\beqa
&& Y_4^{(1,1)} = {{2\,B\,{  b}}\over {{B^2} + {C^2}}}
+ {{\left( {  a}\,{B^2} - A\,B\,{  b} + {  a}\,{C^2} \right) \,
       {  \ln\chi}}\over {{{\left( {B^2} + {C^2} \right) }^{{3\over 2}}}}}
\nonumber \\ 
&& Y_4^{(1,2)} = {{2\,\left( {  a}\,{B^2} - A\,B\,{  b} + {  a}\,{C^2} 
\right) }\over 
     {\left( {A^2} - {B^2} - {C^2} \right) \, +\left( {B^2} + {C^2} \right) }}
+ {{B\,{  b}\,{  \ln \chi}}\over 
     {{{\left( {B^2} + {C^2} \right) }^{{3\over 2}}}}}
\nonumber \\ 
&& Y_4^{(2,1)} = {{{  b}\,\left( 4\,{  a}\,{B^3} - 2\,A\,{B^2}\,{  b} + 
         4\,{  a}\,B\,{C^2} + A\,{  b}\,{C^2} \right) }\over 
     {{{\left( {B^2} + {C^2} \right) }^2}}}\nonumber \\
&&\,\,\,\,\,\,\,\,\,\,\,\,\,\,\,\,\,\,\, + \left( 2\,{{{  a}}^2}\,{B^4} - 4\,A\,{  a}\,{B^3}\,{  b} +
   2\,{A^2}\,{B^2}\,{{{  b}}^2} + 4\,{{{  a}}^2}\,{B^2}\,{C^2} \right.
\nonumber \\
&& \,\,\,\,\,\,\,\,\,\,\,\,\,\,\,\,\,\,\,\,\,\,\,\,\,\,\,\,\,\,\,\,
\left. - 4\,A\,{  a}\,B\,{  b}\,{C^2} - {A^2}\,{{{  b}}^2}\,{C^2} +
   {B^2}\,{{{  b}}^2}\,{C^2} + 2\,{{{  a}}^2}\,{C^4} +
   {{{ b}}^2}\,{C^4} \right) \nonumber \\
&&\,\,\,\,\,\,\,\,\,\, \,\,\,\,\,\,\,\,\,\,\,\,\,\,\,\,\,\,\,
\times {1\over 2 (B^2 + C^2)^{5/2}}\ln \chi
\nonumber \\
&& \chi= {A + \sqrt{B^2 + C^2}\over A - \sqrt{B^2 + C^2}}, 
\eeqa

and similar expressions for the remaining ones. 

The expression of $I_n^{(k,l)}$ in the cases where collinear 
singularities appear, is obtained by comparing the angular integral in the 
imaginary part 
(the s-channel cut) of the 
box diagram
to the imaginary part of the Feynman parameterization of the same diagram 
\cite{thomas}.


\newpage
\section*{Figures}
\vfill
\begin{figure}[bh]
\centerline{\epsfbox{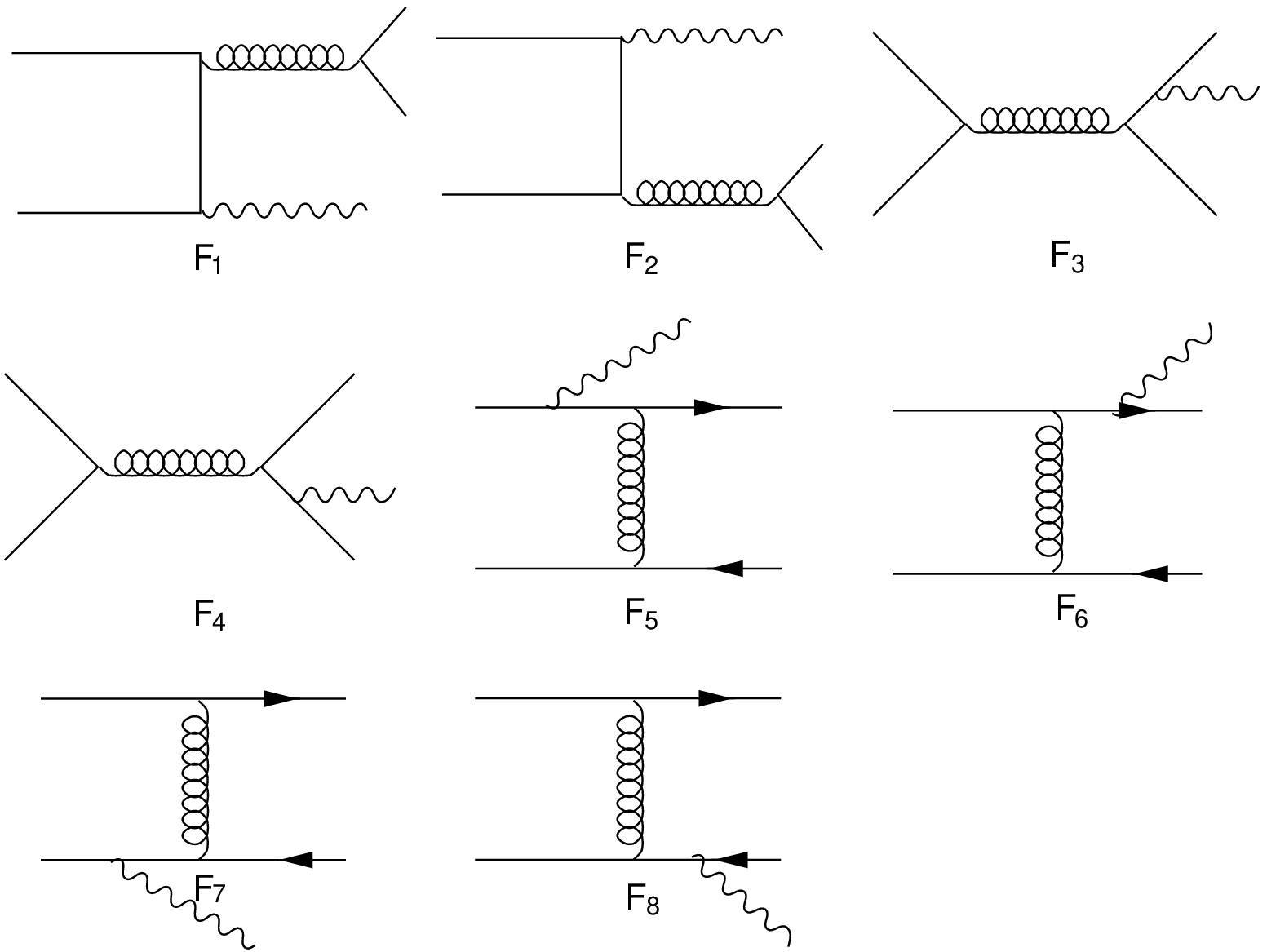}}
\caption{Diagrams which contribute to the process
$q+\bar{q}\rightarrow \gamma^* +q+\bar{q}$}
\end{figure}
\vfill
\begin{figure}
\centerline{\epsfbox{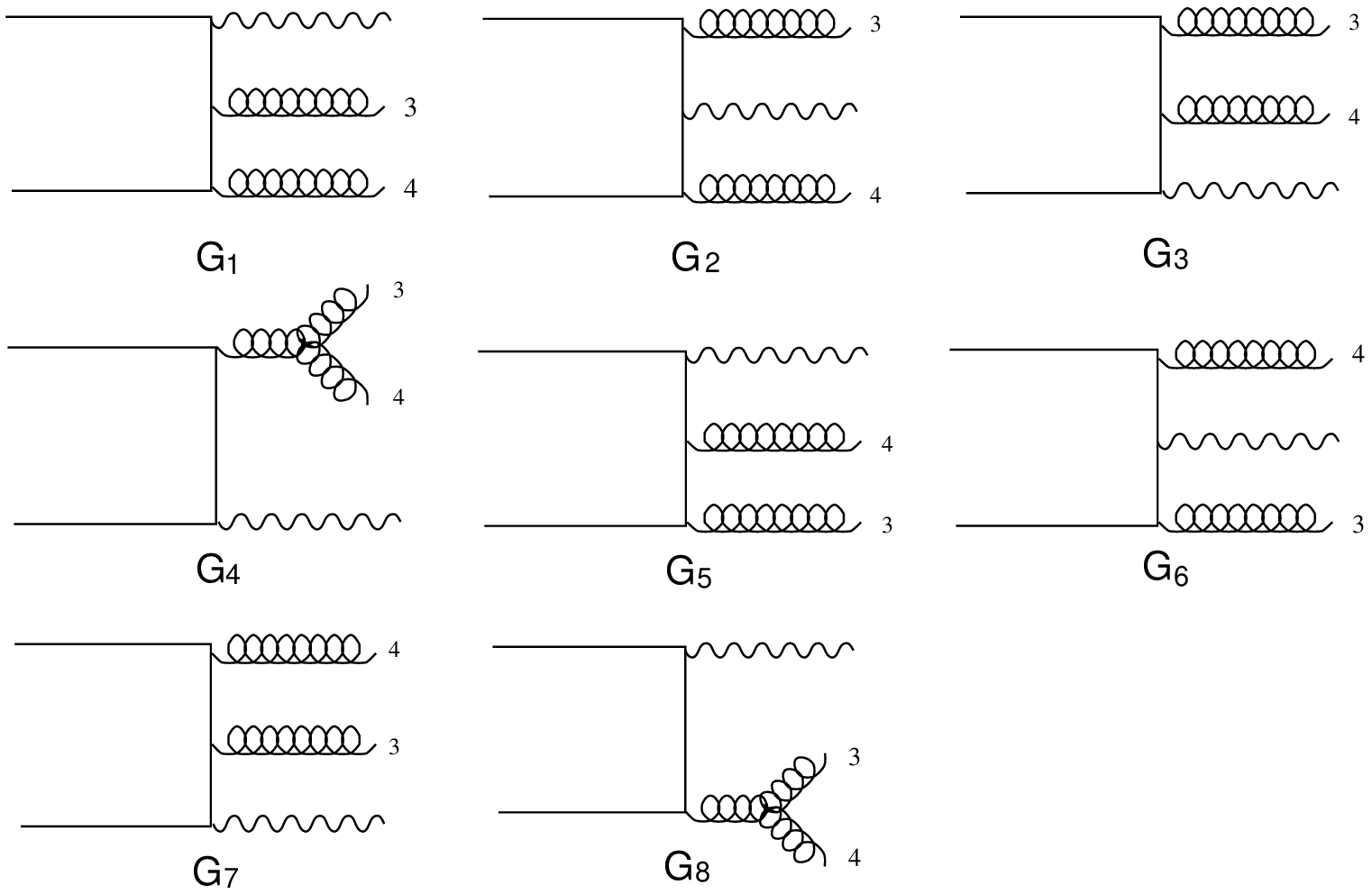}}
\caption{Diagrams which contribute to the process
$q+\bar{q}\rightarrow \gamma^* +G+G$}
\end{figure}
\begin{figure}
\centerline{\epsfbox{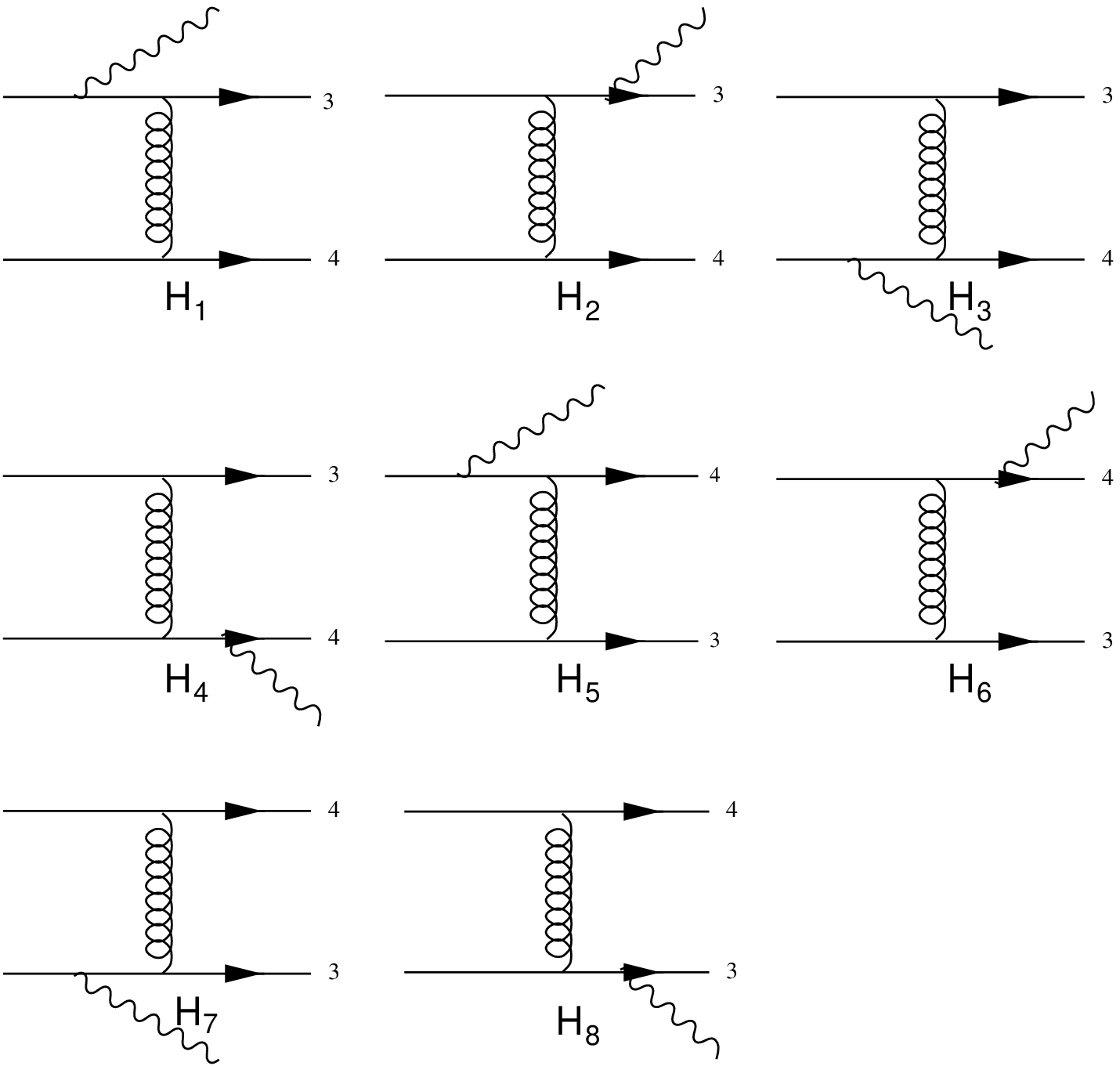}}
\caption{Diagrams which contribute to the process
$q+q\rightarrow \gamma^* +q+q$}
\end{figure}
\begin{figure}
\centerline{\epsfbox{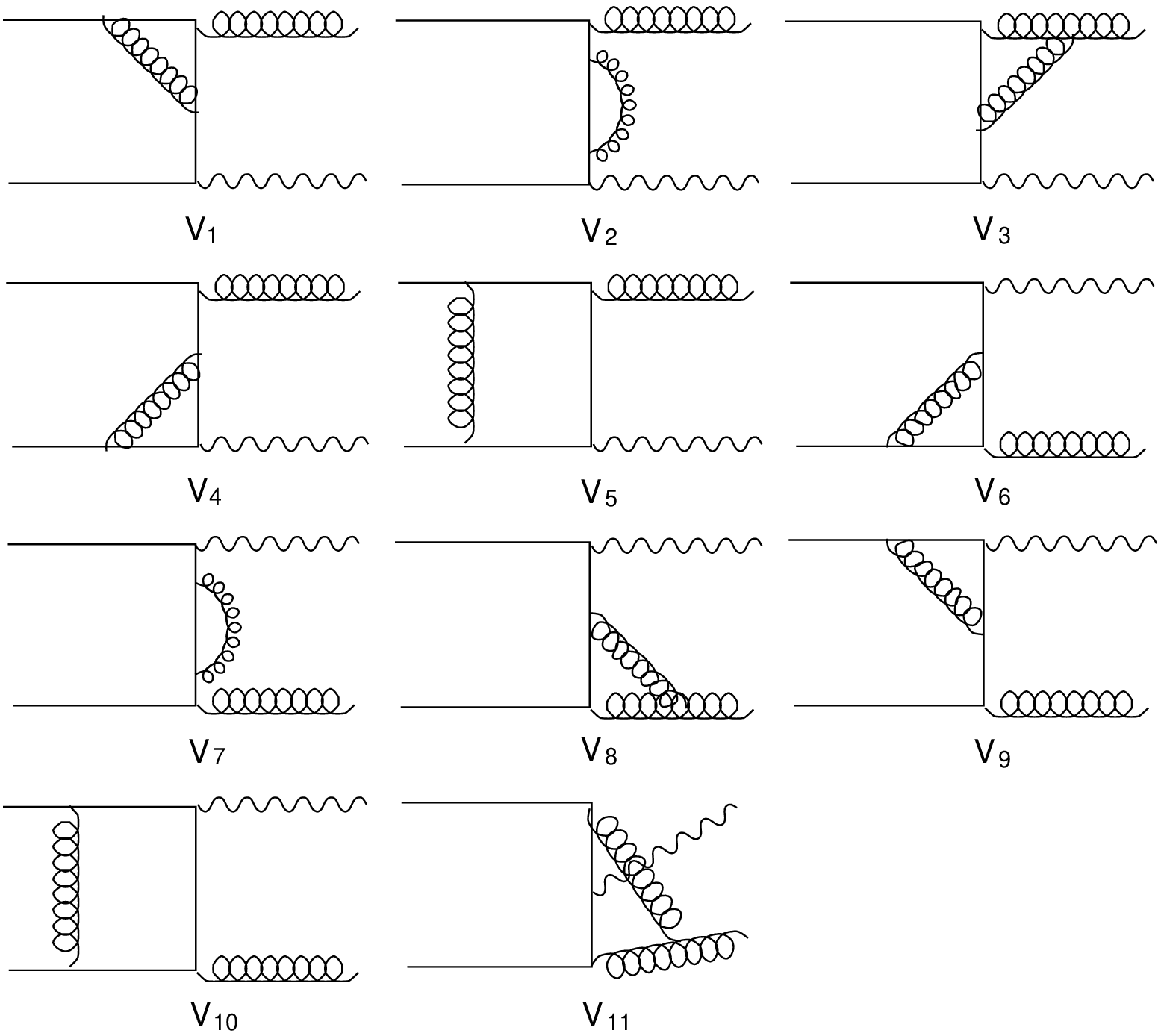}}
\caption{Diagrams which contribute to the process
$q+\bar{q}\rightarrow \gamma^* +G$}
\end{figure}
\begin{figure}
\centerline{\epsfbox{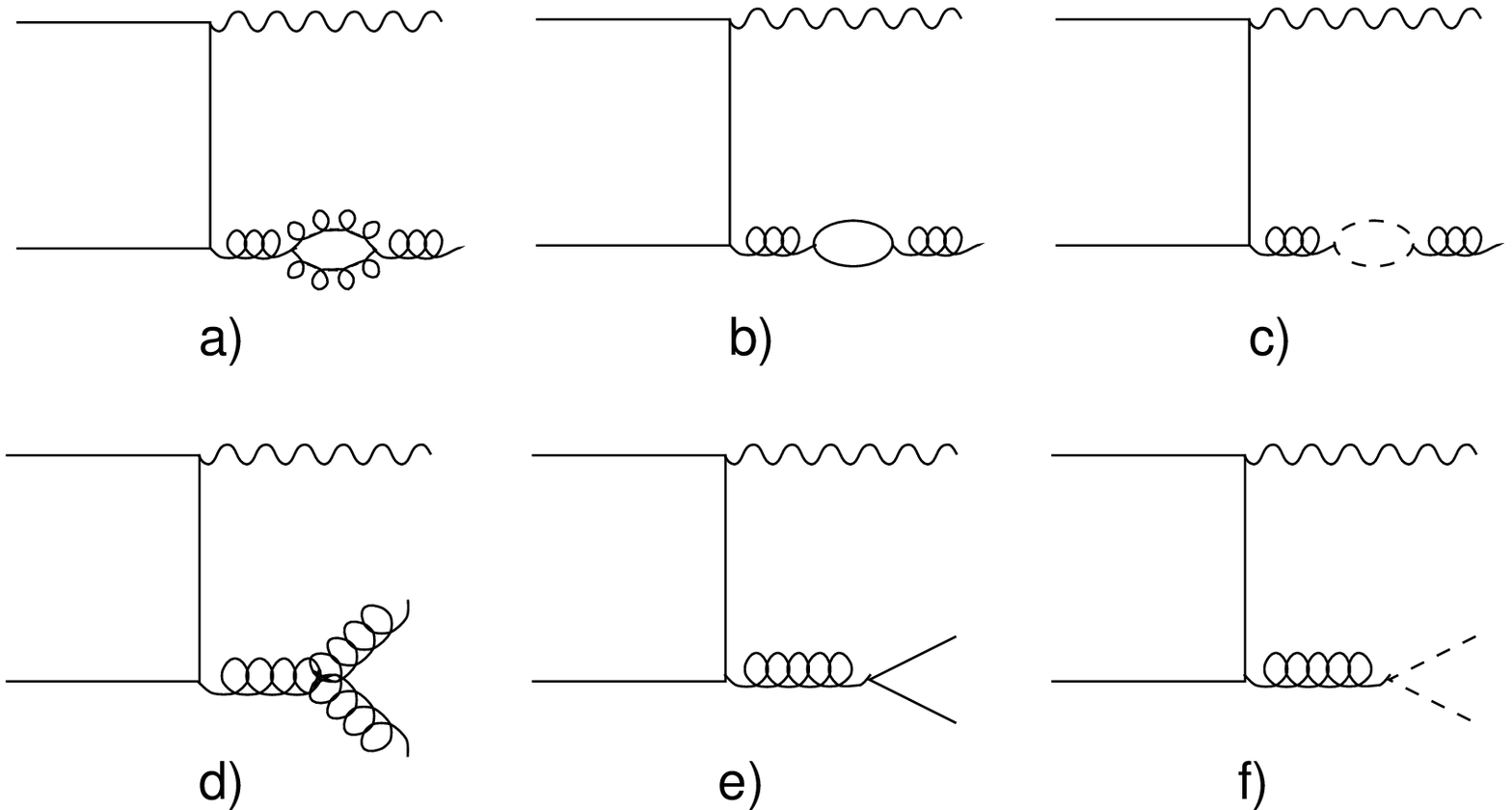}}
\caption{Diagrams which contribute to the process
$q+\bar{q}\rightarrow \gamma^* +G$}
\end{figure}

\end{document}